\documentclass[sigconf]{acmart}
\usepackage{tikz} 
\usepackage{amsmath}
\usepackage{graphicx}

\AtBeginDocument{%
  \providecommand\BibTeX{{%
    \normalfont B\kern-0.5em{\scshape i\kern-0.25em b}\kern-0.8em\TeX}}}

\usepackage{amssymb}
\usepackage{amsfonts}
\usepackage{tabularx,booktabs}
\usepackage{multirow}
\usepackage{caption}
\usepackage{subcaption}
\usepackage{enumitem} 

\usepackage{todonotes}
\usepackage{cleveref}
\usepackage{comment}
\usepackage{xspace}

\copyrightyear{2024}
\acmYear{2024}
\setcopyright{acmlicensed}\acmConference[CHI '24]{Proceedings of the CHI Conference on Human Factors in Computing Systems}{May 11--16, 2024}{Honolulu, HI, USA}
\acmBooktitle{Proceedings of the CHI Conference on Human Factors in Computing Systems (CHI '24), May 11--16, 2024, Honolulu, HI, USA}
\acmDOI{10.1145/3613904.3642033}
\acmISBN{979-8-4007-0330-0/24/05}

\renewcommand{\paragraph}[1]{\vspace*{6pt}\noindent\textbf{#1}\;}

\newcommand*\circled[1]{\tikz[baseline=(char.base)]{
            \node[shape=circle,draw,inner sep=1.25pt] (char) {#1};}} 

\newcommand{\secref}[1]{Section~\ref{#1}}
\newcommand{\figref}[1]{Figure~\ref{#1}}
\newcommand{\tabref}[1]{Table~\ref{#1}}
\newcommand{\apref}[1]{Appendix~\ref{#1}}

\newcolumntype{L}{>{\raggedright\arraybackslash}p} 
\newcolumntype{C}{>{\centering\arraybackslash}p} 
\newcolumntype{R}{>{\raggedleft\arraybackslash}p} 

\newcommand{\fixme}[1]{\ifnum\authnote=1{\textcolor{red}{[FIXME: #1]}}\fi}
\newcommand{\better}[1]{\ifnum\authnote=1{\textcolor{violet}{[BetterWord: #1]}}\fi}
\renewcommand{\todo}[1]{\ifnum\authnote=1{\textcolor{red}{[TODO: #1]}}\fi}

\newcommand{\ulinethick}[1]{%
    \tikz[baseline=(todotted.base)]{
        \node[inner sep=1pt,outer sep=0pt] (todotted) {#1};
        \draw[thick] (todotted.south west) -- (todotted.south east);
    }%
}%
\newcommand{\ulinetwo}[1]{%
    \tikz[baseline=(todotted.base)]{
        \node[inner sep=1pt,outer sep=0pt] (todotted) {#1};
        \draw[double] (todotted.south west) -- (todotted.south east);
    }%
}%

\newcommand{\udotthick}[1]{%
    \tikz[baseline=(todotted.base)]{
        \node[inner sep=1pt,outer sep=0pt] (todotted) {#1};
        \draw[dotted, thick] (todotted.south west) -- (todotted.south east);
    }%
}%

\newcommand{\udensdot}[1]{%
    \tikz[baseline=(todotted.base)]{
        \node[inner sep=1pt,outer sep=0pt] (todotted) {#1};
        \draw[densely dotted, thick] (todotted.south west) -- (todotted.south east);
    }%
}%

\newcommand{\udashthick}[1]{%
    \tikz[baseline=(todotted.base)]{
        \node[inner sep=1pt,outer sep=0pt] (todotted) {#1};
        \draw[dashed, thick, double] (todotted.south west) -- (todotted.south east);
    }%
}%

\newcommand{\udensdash}[1]{%
    \tikz[baseline=(todotted.base)]{
        \node[inner sep=1pt,outer sep=0pt] (todotted) {#1};
        \draw[densely dashed] (todotted.south west) -- (todotted.south east);
    }%
}%

\newcommand{\uloosdash}[1]{%
    \tikz[baseline=(todotted.base)]{
        \node[inner sep=1pt,outer sep=0pt] (todotted) {#1};
        \draw[loosely dashed] (todotted.south west) -- (todotted.south east);
    }%
}%

\newcommand{\vio}[1]{\textcolor{violet}{\ulinethick{#1}}}
\newcommand{\tea}[1]{\textcolor{teal}{\udashthick{#1}}}
\newcommand{\pur}[1]{\textcolor{purple}{\udensdash{#1}}}
\newcommand{\cya}[1]{\textcolor{cyan}{\udotthick{#1}}}
\newcommand{\ora}[1]{\textcolor{orange}{\ulinetwo{#1}}}

\newcommand{\blu}[1]{\textcolor{blue}{\udensdot{#1}}}

\newcommand{\maj}[1]{\textcolor{magenta}{\uloosdash{#1}}}

\newcommand{\edit}{\textcolor{black}}

\newcommand{\setdupe}{{\textcolor{black}{513}}\xspace} 
\newcommand{\set}{{\textcolor{black}{464}}\xspace} 

\newcommand{\target}{{\textcolor{black}{complainant}}\xspace} 
\newcommand{\targetter}{{\textcolor{black}{abuser}}\xspace} 

\newtoggle{shownotes}
\toggletrue{shownotes}
\iftoggle{shownotes}{%
    \newcommand{\knote}[1]{{\color{magenta}[ {{#1}} -- Kevin]}}
    \newcommand{\ronote}[1]{{\color{red}[ {{#1}} -- Rosie]}}
    \newcommand{\anote}[1]{{\color{purple}[ {{#1}} -- Arka]}}
    \newcommand{\vnote}[1]{{\color{green}[ {{#1}} -- Vineeth]}}
    \newcommand{\jnote}[1]{{\color{orange}[ {{#1}} -- Jessica]}}
}{%
    \newcommand{\knote}[1]{\xspace}
    \newcommand{\ronote}[1]{\xspace}
    \newcommand{\anote}[1]{\xspace}
    \newcommand{\vnote}[1]{\xspace}
    \newcommand{\jnote}[1]{\xspace}
}

\newtoggle{redactorg}
\toggletrue{redactorg}
\iftoggle{redactorg}{%
    \newcommand{\redact}[1]{{{[redacted organization]}}}
}{%
    \newcommand{\redact}[1]{{{#1}}}
}            

\hypersetup{
	colorlinks   = true, 
	urlcolor     = black, 
	urlbordercolor=black, 
	linkcolor    = blue, 
	linkbordercolor = white, 
	citecolor    = blue, 
	citebordercolor = white, 
	breaklinks   = true, 
}
\makeatletter 
\Hy@AtBeginDocument{
	\def\@pdfborder{0 0 1} 
	\def\@pdfborderstyle{/S/U/W 0.5} 
}
\makeatother
\begin{document}

\title{Shortchanged: Uncovering and Analyzing Intimate Partner Financial Abuse in Consumer Complaints}



\author{Arkaprabha Bhattacharya}
\orcid{0009-0009-4585-0280}
\affiliation{%
  \institution{JPMorgan Chase}
  \city{New York}
  \state{NY}
  \country{USA}
}

\author{Kevin Lee}
\orcid{0000-0001-5416-4826}
\affiliation{%
  \institution{JPMorgan Chase}
  \city{New York}
  \state{NY}
  \country{USA}
}

\author{Vineeth Ravi}
\orcid{0009-0003-0424-5531}
\affiliation{%
  \institution{JPMorgan Chase}
  \city{New York}
  \state{NY}
  \country{USA}
}

\author{Jessica Staddon}
\orcid{0009-0001-9361-9413}
\affiliation{%
  \institution{JPMorgan Chase}
  \city{Palo Alto}
  \state{CA}
  \country{USA}
}

\author{Rosanna Bellini}
\orcid{0000-0002-2223-2801}
\affiliation{%
  \institution{Cornell University}
  \city{New York}
  \state{NY}
  \country{USA}
}
\renewcommand{\shortauthors}{Bhattacharya et al.}

\begin{abstract}
\noindent Digital financial services can introduce new digital-safety risks for users, particularly survivors of intimate partner financial abuse (IPFA).
\edit{To offer improved support for such users, a comprehensive understanding of their support needs and the barriers they face to redress by financial institutions is essential.} 
Drawing from a dataset of 2.7 million customer complaints, we implement a bespoke workflow that utilizes language-modeling techniques and expert human review to identify complaints describing IPFA. 
Our mixed-method analysis provides insight into the most common digital financial products involved in these attacks, \edit{and the barriers consumers report encountering when doing so.} 
Our contributions are twofold; we offer the first human-labeled dataset for this overlooked harm and provide practical implications for \edit{technical} practice, research, and design for better supporting and protecting survivors of IPFA.

\end{abstract}




\maketitle
%
\section{Introduction}
\label{sec:intro}
\noindent Financial abuse --- the control of access to, and maintenance of, financial resources \cite{sharp2015review} --- is a devastating form of intimate partner violence (IPV) that severely impacts the mental, physical, and spiritual wellbeing of those targeted.
Such individuals are marginalized, highly vulnerable \cite{postmus_building_2023, postmus2012understanding, sharp2015review}, and subject to substantive digital-safety risks from a targeted adversary \cite{warford_sok_2022}. 
An \targetter may know 
confidential information about a survivor and possess complex social goals that go beyond financial gain.
In these contexts, consumer-facing banking applications can facilitate the surveillance of a \target's expenditure through online interfaces \cite{bellini2023digitalsafety} or monitoring alerts \cite{ellie_butt_know_2020}, while coercive uses of authorized user status can enable an abuser to fraudulently make purchases or coerced debt \cite{Bellini2023paying}.

Digital financial products and services can exacerbate existing harms \cite{bellini2023digitalsafety, Bellini2023paying, ellie_butt_know_2020}, since they are rarely designed with digital-safety concerns in mind.
While consumer-facing technologies play a growing role in these contexts \cite{ellie_butt_know_2020, Bellini2023paying}, identifying attacks is nevertheless fraught with difficulties \cite{tseng_digital_2021}.
Many technology-enabled attacks are reported retroactively, often requiring survivors to make inferences around how an attack was conducted or the vulnerabilities that made them possible \cite{freed_digital_2017}.
It is also extremely difficult to reach survivors of financial abuse due to their marginalized and vulnerable status, necessitating extensive care and attention in research endeavors \cite{tseng2022care}.

Financial institutions, including banks, credit bureaus, and insurance companies, have been recognized as crucial safeguarding environments for the financial well-being of survivors \cite{scott_financial_2023, barros_pena_pick_2021, consumer_financial_protection_bureau_cfpb_2023}. 
However, to do so, such institutions need to be equipped with an awareness of the challenges a survivor may face --- particularly as more survivors may reach out to such organizations for complaints about consumer-facing products \cite{zou_role_2021}.
While prior work has described the attacks experienced by complainants currently receiving support via IPV services \cite{freed2018stalker, woodlock2017abuse}, and how abusers may craft such attacks online \cite{tseng2020tools, bellini2021so}, we present a detailed view of how these concerns about digital products may be presented to financial institutions firsthand.
This study examines how consumer-authored narratives on intimate partner-perpetrated financial abuse can provide insights into:

\begin{enumerate}[leftmargin=3em, label=\bfseries RQ\arabic*:]
\vspace{-0.8em}
\setlength\itemsep{0em}
    \item How might computational text analysis help to identify financial abuse between intimate partners in online consumer complaints?
    \item Which digital consumer-facing financial products and technology-enabled financial attacks are prominently represented in such complaints?
    \item What barriers to service do consumers report encountering when attempting to resolve concerns around technology-enabled financial abuse?
\end{enumerate}

We collaborated with industry and academic experts on IPV and customer safety to create a tailored workflow \edit{for collecting relevant complaints}. 
Our workflow combines pre-trained language models with careful human review to identify instances of financial abuse in text-based consumer complaints. 
Using the vast Consumer Financial Protection Bureau (CFPB) Consumer Complaint Database --- totaling over 2.7 million entries --- and our workflow, we generated a specialized dataset of \setdupe consumer accounts reporting financial abuse to financial institutions. 
Utilizing Framework Analysis \cite{ritchie_qualitative_2013} and Critical Discourse Analysis \cite{blommaert_critical_2000, kress_critical_1990, fairclough_critical_2012}, we characterize when survivors reach out to the CFPB for help, which digital products they cite as a cause for concern, what barriers they encounter while doing so, and how much they report losing financially.
Their path through the complaint process at the financial institution is lengthy and complex, with reported challenges in policy design, the need for digital evidence collection, and considerable digital-safety risks.

\edit{Our findings offer insight into how to improve the safety of digital financial products for survivors of financial abuse.
To mitigate financial attacks, we offer suggestions for how to implement safety checkups and regular system audits in financial products as promising strategies in this area.
Our work also highlights the importance of considering new approaches to evidence gathering and reporting approaches when reaching out to institutions for help, and better align with the needs of vulnerable customers.
We conclude by highlighting new research directions in technology-enabled financial abuse that are especially poignant in light of the growing adoption of digital financial products.}

%
\section{Background and Related Work}
\label{sec:relwork}
\noindent 
In this study, our focus is on targeted digital financial attacks in situations involving intimate partner violence.
These attacks, carried out by an abuser, can result in the loss of wealth, property, or other financial benefits for a survivor \cite{stylianou2018economic, sharp2015review}, for a range of social, or financial motives \cite{freed_digital_2017, Bellini2023paying}.
We refer to these actions as \textit{intimate partner financial abuse} (IPFA), which may also branch into elder financial abuse \cite{fraga2022elder, latulipe_unofficial_2022} if involving older adults \edit{(see \apref{sec:app-definitions} for differentiation)}.
To align with current best practice, we employ the term \textit{survivor} to honor an individual's strength and resilience in the face of adversity, and use \textit{abuser} to identify the direction of abuse \cite{surviving_economic_abuse_conversation_2021, freed2018stalker, freed_digital_2017}.

Digital technologies play a substantive role in enabling abusers to coerce, control, harass and otherwise harm their current or former intimate survivors.
Recent work has shown that these actions extend to financial and economic sectors, from surveillance of financial expenditure \cite{chatterjee_spyware_2018, tseng2020tools}, to permitting unauthorized entry into smartphone applications \cite{bellini2023digitalsafety}, to harassing messages sent in payment memos \cite{ellie_butt_know_2020}.
In their in-depth review of survivor accounts of tech-enabled abuse, Bellini \cite{Bellini2023paying} identified a range of technical attacks across a variety of technical products, including credit accounts, shared banking infrastructures, and online businesses.
Thus, abusers defy the conventional threat models of consumer technologies by exerting physical control over a survivor's devices and exploiting use of private financial information.
While some scholars have identified that flexible, proportionate, and consentful design could mitigate certain negative effects of some consumer technologies \cite{barros_pena_pick_2021}, these designs are unfortunately still in their early stages and not widely embraced by mainstream services.
\edit{Additionally, there is insufficient knowledge about which financial products or features are prone to targeting \cite{bellini2023digitalsafety, Bellini2023paying}, hindering the ability for financial service providers to make impactful changes for users.}

Many human-computer interaction (HCI) and computer security researchers have investigated interventions that work to support survivors of technology-facilitated abuse.
Approaches to clinical computer security that pair survivors with privacy and security experts have shown promise in protecting survivors' devices and preventing their digital footprints from being targeted \cite{havron_clinical_2019, freed_is_2019}.
In such technology clinics, experts are able to provide in-depth insight into the types of digital technologies that incur unique risks to survivors \cite{tseng2022care}, while also helping to professionalize such services.
Zou et al. \cite{zou_role_2021} also provide a welcome focus on customer support agents at computer security companies who may provide assistance to survivors during or in the immediate aftermath of a technical attack.
There remains a paucity of research on how survivors of technology abuse, particularly in financial contexts, may reach out in ways beyond expected customer service channels \cite{zou_role_2021}, or direct referrals to specialized technology services \cite{freed_is_2019, havron_clinical_2019, tseng2022care}.
This is in spite of the increasing recognition that financial institutions and similar organizations can serve as powerful social and political influencers due to their close connections with customers \cite{phelan_financial_2021, scott_financial_2023, branicki_corporate_2023, kemp2005elder, barros_pena_pick_2021}.
\edit{Thus, obtaining deeper insights into technology-enabled financial abuse could offer significant ways forward for intervention and technical implementation for financial institutions and designers} \cite{karam_intimate_2023, branicki_corporate_2023, scott_financial_2023}.

Obtaining such first-hand insights into the financial actions of vulnerable service users \edit{to inform our gaps in knowledge about IPFA and digital systems} is, understandably, challenging.
For many consumers, the mere discussion of finances can be culturally taboo \cite{barros_pena_financial_2021}; out of fear of manipulation by others \cite{Alsoubai2022friends}, or by experiencing embarrassment or shame \cite{vitak_i_2018}.
Financial institutions also go to great lengths to maintain customer confidentiality about information relating to their finances to both prevent fraud and protect against the disclosure of proprietary information.
\par Deep learning and natural language processing have been used to parse and analyze large datasets, which have enabled the inference of otherwise inaccessible characteristics  \cite{saxena_unpacking_2022, antoniak_riveter_2023, razi_human-centered_2021}.
Scholars have leveraged these approaches to elicit risk factors to vulnerable groups, contextualize taboo topic areas (e.g., end-of-life~\cite{yao_join_2021}, climate change \cite{falkenberg2022growing}), and identify under-recognized forms of socio-technical harms \cite{Alsoubai2022friends, hanley2022happenstance}.
In the context of abuse and hate, Bidirectional Encoder Representations from Transformers (BERT)-based transformer models have been deployed to detect hate speech online \cite{koufakou2020hurtbert, wullach2021fight}, identify descriptions of abuse in patient records \cite{botelle_can_2022}, and detect abusive transactions \cite{leontjeva_detection_2023}.
\edit{However, at the time of writing, we have yet to locate research that has grappled with the challenges of applying such approaches to better contextualize technology-enabled abuse for survivors of IPFA.}

%
\section{Study and Workflow Design}
\noindent Technology-enabled financial abuse by intimate partners is an emergent area of research \cite{bellini2023digitalsafety, Bellini2023paying, ellie_butt_know_2020}, involving the identification of subtle and complex factors that even human agents may struggle to identify.
While many HCI works have shown that machine learning may hold many benefits in addressing under-explored problem spaces, we acknowledge it  \textit{``has no clairvoyant abilities''} \cite{arp2022and}, thus requiring careful reasoning about data, workflow, and derived conclusions to resist negative knock-on-effects \cite{mathur_disordering_2022, salminen_ethics_2020, chancellor_taxonomy_2019, chancellor_towards_2022}.
We aim to address this challenge head-on \edit{through our exploration of \textbf{RQ1} by our study design,} which involves synthesizing natural language processing with careful manual review with experts; a \edit{technique} that has seen promise in other areas  \cite{ratner2019accelerating, harkous2022hark, nema2022analyzing}.
\par In this section, we describe our \textit{study context}, and our approach to \textit{data collection and cleaning}.
We then describe the \textit{workflow to identify IPFA complaints}, accompanied by an \textit{assessment of its effectiveness} through techniques designed for explaining machine learning model decisions. We follow up to this with an overview of our dataset and analysis, before concluding with a reflection on \textit{ethical considerations}.


\paragraph{Study context: public consumer complaints.}
\label{sec:cfpb}
\noindent We investigate IPFA in the context of unstructured complaints written by consumers and submitted to the Consumer Financial Protection Bureau (CFPB), a United States \edit{(U.S.)} government agency responsible for consumer protection in the financial sector.
Since the CFPB's inception in 2011 through the Dodd-Frank Act\footnote{12 U.S. Code § 5491 - Establishment of the Bureau of Consumer Financial Protection}, the U.S. Congress has directed the bureau to collect and monitor complaints from consumers about financial products and services from over 6,100 financial companies in order to promote transparency and fairness for consumer-facing financial products and services (e.g., credit reporting and mortgage lending).\footnote{CFPB supervisory powers cover banks, thrifts, and credit unions with assets over \$10 billion, along with non-bank mortgage originators, payday lenders, and private student lenders. Recently, these powers expanded to include consumer reporting, student loan servicing, international money transfer, automobile finances, and consumer debt collection.} 
\par The CFPB handles its primary duty to collect, investigate, and resolve consumer complaints via a toll-free hotline or through a secure online portal for companies.
Consumers may submit complaints to the CFPB, either before or after contacting their financial institution(s), which includes financial service providers and credit bureaus. 
If a consumer disagrees with the financial institution's resolution to their concern, they can file a complaint with the CFPB. 
Complaints include free-text narratives (subject to a 10,000 character limit) and can categorize their concerns via selections from pre-populated menus. 
The CFPB forwards verified complaints to financial organizations, which must then respond per service level agreements and legal guidance ~\cite{foohey2017calling}.

Over 50\% of complainants consent to making their complaint publicly available once personally identifiable information has been redacted \cite{narrative_scrubbing}.
The CFPB publishes over 10,000 complaints monthly that have undergone redaction \cite{consumer_financial_protection_bureau_cfpb_2023}. 
Complaints in this dataset have been the center of past analyses, showcasing complaint trends, regional submissions, company response rates, and latent topics \cite{ayres2013skeletons, bastani2019latent}, thus indicating the value of this source for understanding financial abuse.
\par \edit{Survivors' complaints about specific financial products with respect to privacy, security, and safety concerns, are generally inaccessible for external research purposes because they often require access to proprietary or sensitive information.
Alternatively, existing studies that elicit such findings are performed after a survivor has received services --- such as Bellini's \cite{Bellini2023paying} retroactive case study review of service users, which does not cover all survivors \cite{sharp2015review}.
Our distinctive approach} --- that focuses on survivors as \textit{consumers} thus entitled to consumer rights --- affords us the opportunity to harness real-world data and scrutinize emerging trends and critical issues.
\edit{Further, this approach hopes to enhance the existing body of knowledge in HCI on technology-enabled abuse \cite{freed_digital_2017, freed2018stalker, bellini2021so} by concentrating on grievances specifically related to the digital financial products and services implicated by IPFA.} 

%
\paragraph{Data collection and cleaning.}
\label{sec:data_tools}
We downloaded a working dataset of 2,760,540 complaints (2.2 GB) from the CFPB website in June 2022.
Each complaint is represented as a data entry of 18 data fields, including complaint date, product description, customer narratives, and company (\Cref{tab:cfpb-database-schema}, \Cref{sec:app-cfpb-workflow}).
Our analysis is centered on the customer complaint narratives (CCNs), which are customer-authored free-form text without character limits.
Complaints with CCNs averaged 1,043 words, with significant variability (SD 1,276 characters; \Cref{fig:whole_corpus_lengths}, \Cref{sec:app-cfpb-workflow}). 
\par To focus on consumer-identified financial abuse complaints, we excluded 1,786,680 complaints without a CCN. 
We then identified and removed same-day, same-time duplicates (indicative of a technical error), while retaining duplicates that we judged to be customer-authored resubmissions (i.e., complaints of identical content sent to different financial institutions)
As such, our total dataset contains 973,860 complaints. \edit{All tokenization, cleaning, and keyword searches in our workflow were performed with the use of the \texttt{spacy} library (version numbers for all tools are in \Cref{tab:app-tools-versions}, \Cref{sec:app-tools})}. As we are unable to establish ground truth in the authorship of each complaint, we refer to the user who wrote the complaint as the \textit{complainant}, and retain the terminology of \targetter.

\begin{figure*}[t]
    \centering
    \includegraphics[width=\textwidth,trim={4.5mm 0 0 0},clip]{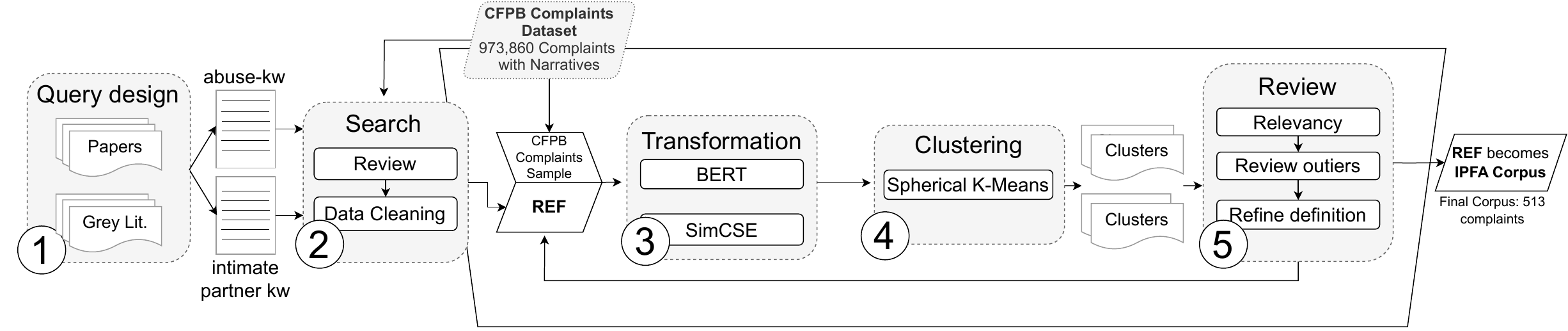}
    \caption{\small The IPFA complaint collection workflow involves a reference set (\textit{ref}) of known IPFA examples and unlabelled CFPB complaints containing an intimate partner keyword. Text embeddings are created using a sentence transformer model, and these embeddings are then clustered. The clusters with the most reference set complaints are manually reviewed to identify additional IPFA examples, which are then added to the reference set. This process is iterated with new IP complaints. \edit{When the desired iterations are complete, \textit{ref} becomes the final collected dataset.}}
    \label{fig:model_pipeline}
    \Description[Describing the IPFA complaint collection workflow]{The IPFA complaint collection workflow as depicted involves a reference set (labeled ref) of known IPFA examples and unlabelled CFPB complaints containing an IP keyword. Text embeddings are created using a sentence transformer model, and these embeddings are then clustered.The clusters with the most reference set complaints are manually reviewed to identify additional IPFA examples, which are then added to the reference set. This process is iterated with new IP complaints. When the number of desired iterations is complete, REF becomes the final dataset of IPFA complaints.}
\end{figure*}



Survivors of financial abuse in IPV contexts are doubly marginalized; given that money is often a highly taboo discussion topic, and survivors face substantial stigmatization from wider society \cite{warford_sok_2022}. While survivors may feel empowered by sharing personal stories  \cite{tseng_digital_2021}, it may also exhaust an individual already suffering from significant trauma, resulting in additional responsibility on them to share their experiences \cite{Bellini2023paying}. 
We are aware that despite our study motivation to reduce harm to at-risk groups, we did not directly engage survivors of financial abuse to share their personal stories which poses an ethical tension \cite{bellini:online}. After speaking to several experts on both IPV and financial abuse, we were cautious to not directly engage survivors before we learned more about this area \cite{bellini:online}.
We plan to use the insights gained from this work to design a future study that engages with survivors directly.

\paragraph{IPFA identification workflow.}
\label{sec:pipeline}
\noindent To identify this subtle form of financial harm in consumer accounts \cite{bellini2023digitalsafety, Bellini2023paying}, \edit{and help to provide answers to \textbf{RQ1}}, we designed a five-step (\textbf{\circled{1}---\circled{5}}) iterative workflow that synthesizes manual human expert review, natural language processing, and information retrieval techniques (\Cref{fig:model_pipeline}).
First, we carefully designed our query for IPFA cases involving technology abuse \textbf{\circled{1}} via a range of different keywords from prior literature \cite{sharp2015review, Bellini2023paying, postmus_building_2023} to produce two lists of English-language keywords associated with intimate partners (\textit{intimate partner keywords}) and financial abuse (\textit{financial abuse keywords}) (Appendix~\ref{sec:app-query}).
To evaluate the performance of using \textit{intimate partner keywords} for identifying complaints mentioning intimate partners, two authors independently reviewed \edit{a sample of} 100 CCNs \edit{that were flagged if a keyword from \textit{intimate partner keywords} was present, to confirm if a flagged complaint mentioned an intimate partner. 
Both reviewers} were able to resolve all disagreements, yielding a precision of 0.95 and recall of 0.93 \edit{for this set of keywords}. 
\edit{We observed in a small sample of IPFA-relevant complaints (identified by random selection and manual review) that an abuse keyword often appeared in close proximity to a mention of an intimate partner. Using this finding, w}e then enacted a keyword search \textbf{\circled{2}} by searching all 973,860 complaints for cases containing at least one phrase from \textit{intimate partner keywords} and at least one phrase from \textit{financial abuse keywords} within a 10-word proximity to each other.
We refer to this process as \textit{keyword matching with proximity}, which resulted in 1,179 matches. \edit{To further verify the matched complaints and remove matches that did not pertain to IPFA,
two human labellers reviewed the complaints to identify relevance, resulting in an initial reference set (\textit{ref}) of $288$ complaints.}

\begin{figure*}[!htb]
\scriptsize
     \centering
\begin{subfigure}[b]{0.45\textwidth}
    \centering
    \includegraphics[width=\textwidth]{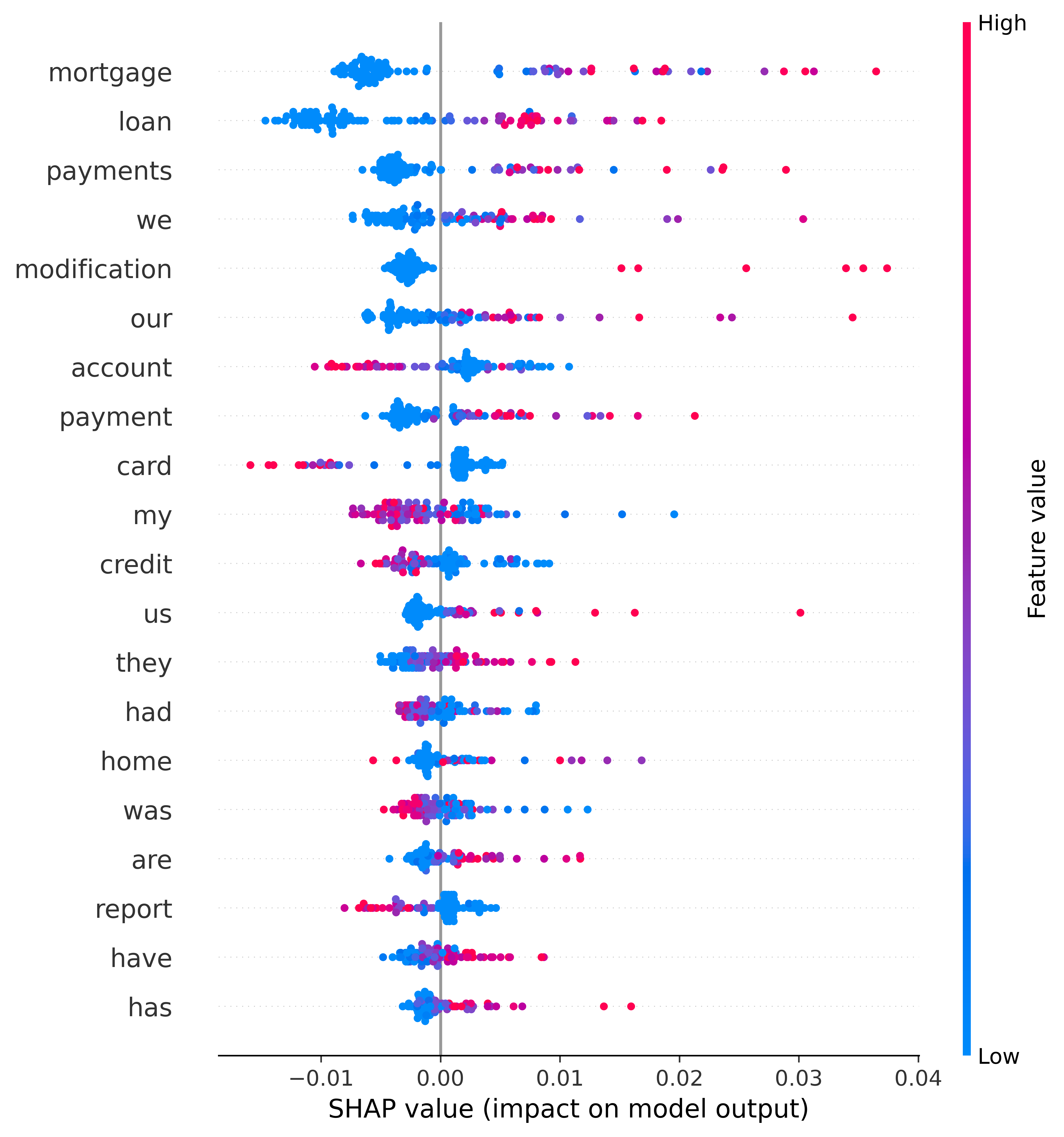}
    \end{subfigure}
     \hfill
\begin{subfigure}[b]{0.45\textwidth}
         \centering
    \includegraphics[width=\textwidth]{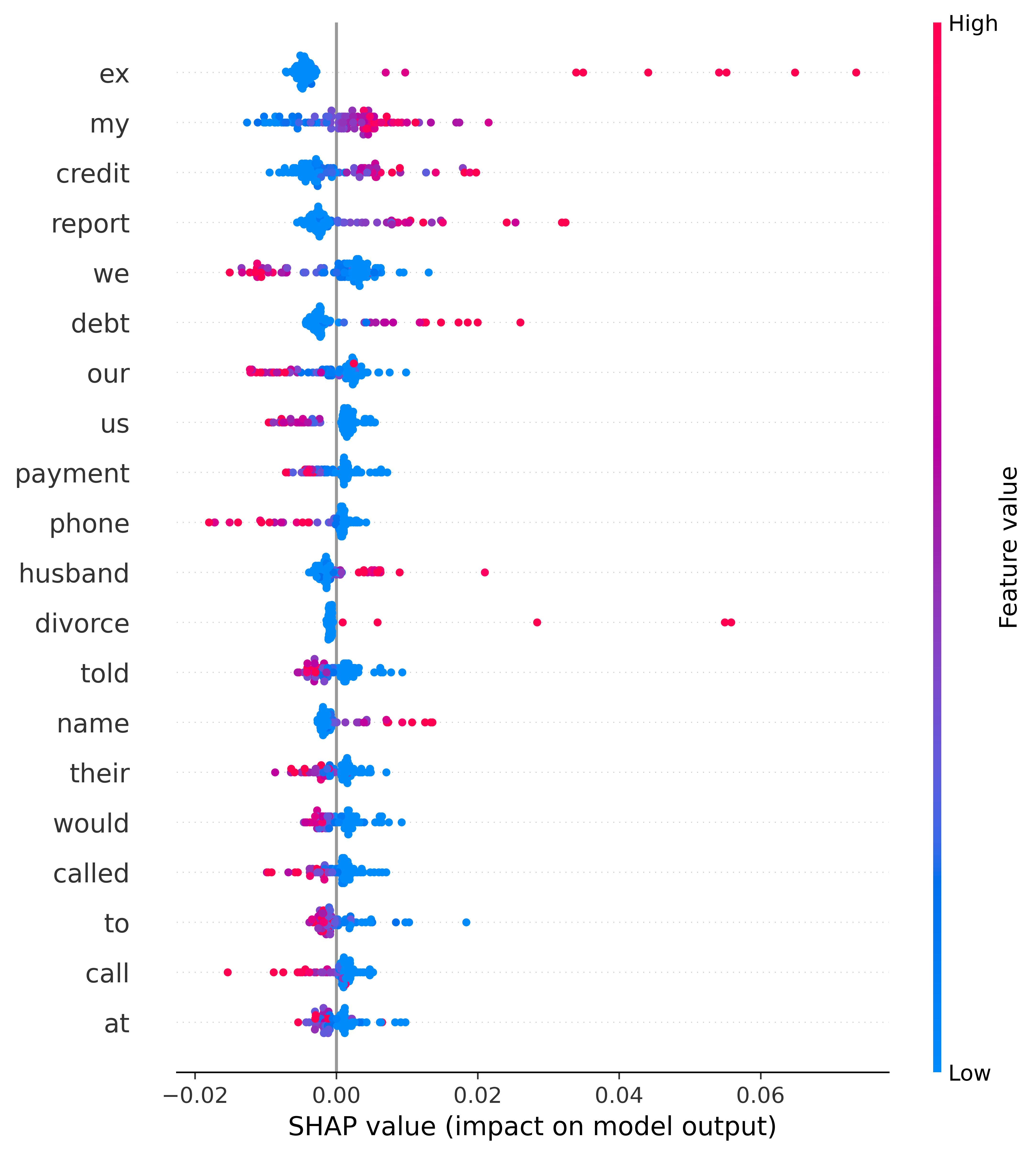}
     \end{subfigure}
     \hfill

\caption{SHAP scores for K-Clusters 1 [Left], and 6 [Right]. The color bar corresponds to the raw values of the variables for each instance. If the variable for a particular word is high, it appears as a red dot, while low variable values appear as blue. \Cref{fig:SHAPs-scores} in \Cref{sec:app-cfpb-workflow} shows SHAP scores for all clusters. }
\label{fig:SHAPs-scores-small}
\Description[SHAP summary scores for K-Clusters.]{SHAP summary scores for K-Clusters from an experiment with cluster 1 [Left], cluster 6 [Center], and cluster 7 [Right] depicted. Cluster 6 and 7's summaries show that inclusive words such as "we" or "our" negatively influence a complaint's classification towards that cluster. Cluster 0's summary shows that this influence has a positive effect for complaints being classified as a part of this cluster.}
\end{figure*}

We used text embeddings (\textbf{\circled{3}}) --- text represented as numerical vectors such that semantically similar text results in vectors that are nearby in the embedding space --- \edit{to represent the identified IPFA complaints from \textbf{\circled{1}} (\textit{ref}) and a sample of new complaint narratives from the original dataset.} We instantiated these text embeddings from sentence transformer models, \textit{bert-large-nli-stsb-mean-tokens} and \textit{sup-simcse-roberta-base} \cite{gao2021simcse, reimers-2019-sentence-bert}. \edit{The models themselves were instantiated with the \texttt{sentence-transformers} and \texttt{simcse} libraries.}

\begin{figure*}[!htb]
\scriptsize
     \centering
     \begin{subfigure}[t]{0.3\textwidth}
         \vskip 0pt
         \centering
         \includegraphics[width=\textwidth, trim={9mm 3mm 22mm 1cm},clip]{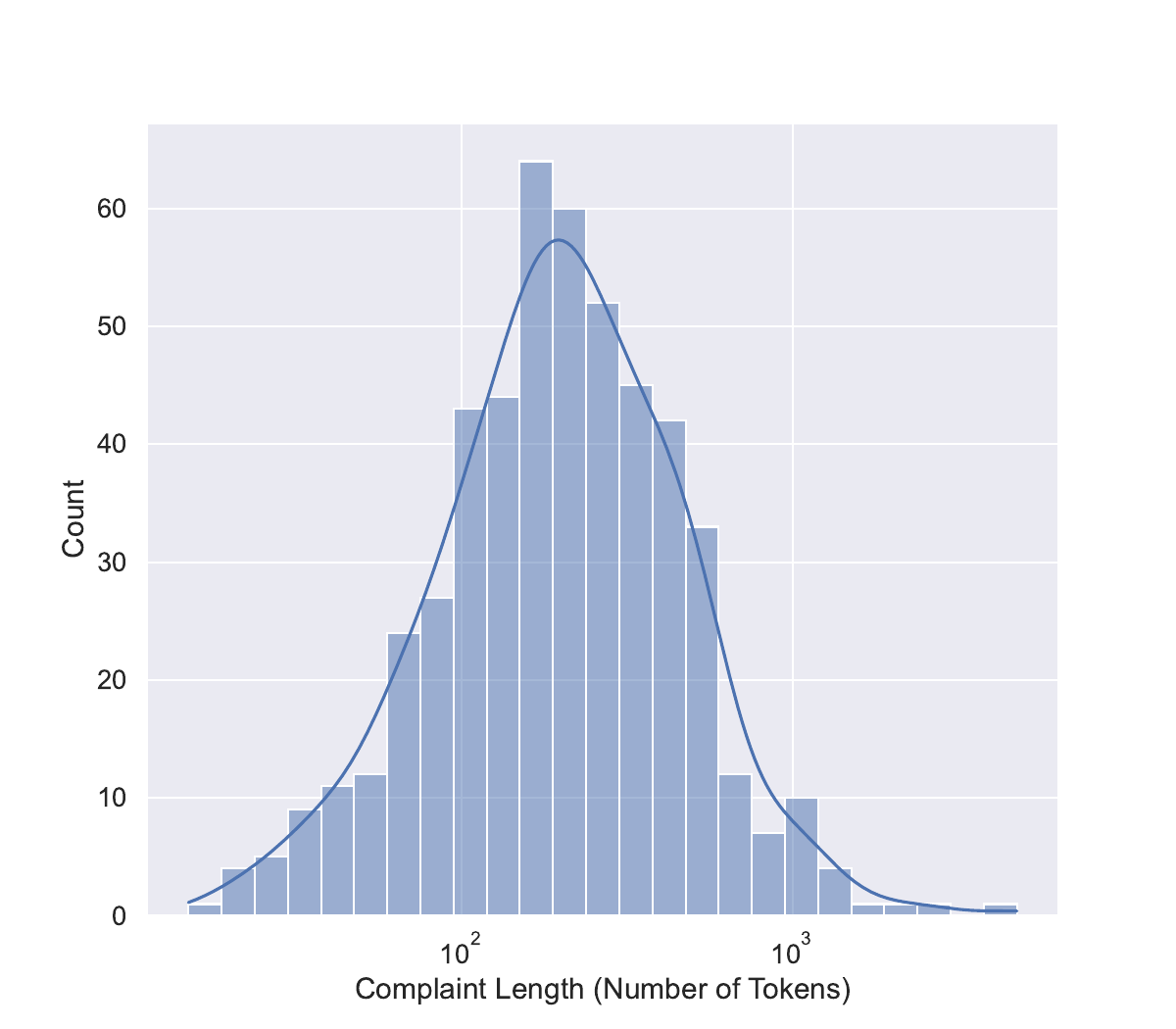}
         \caption{The frequency distribution of complaint lengths (by number of tokens).}
         \label{fig:corpus-lengths}
         \Description[]{A frequency distribution of complaint lengths based on the number of tokens in each complaint, with the average length being 283.5 tokens and the standard deviation being 333.8 tokens.}
     \end{subfigure}
     \hfill
     \begin{subfigure}[t]{0.3\textwidth}
        \vskip 0pt
         \centering
         \includegraphics[width=\textwidth, trim={7mm 5mm 20mm 9mm},clip]{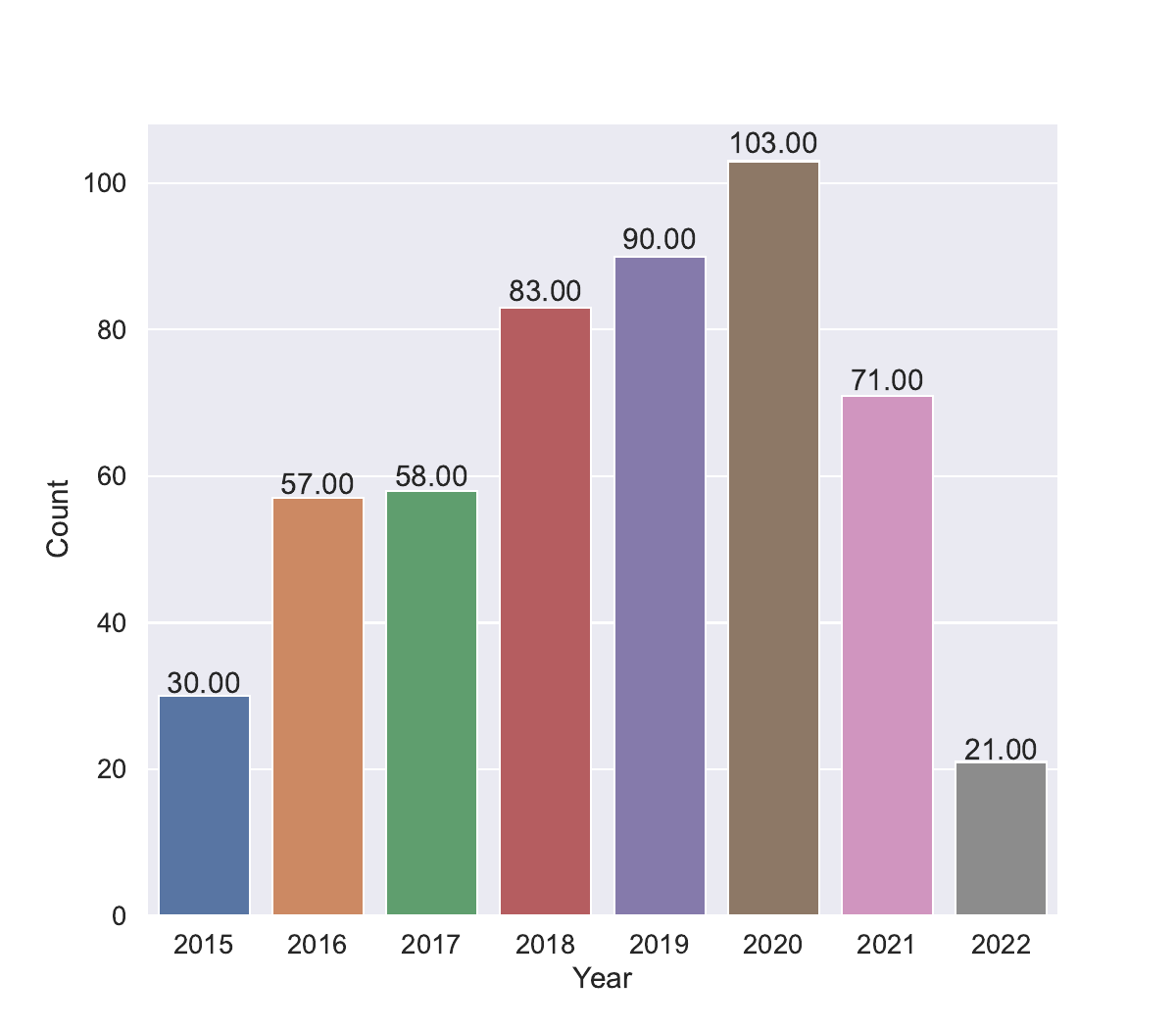}
         \caption{The frequency distribution of the number of complaints submitted per year (2015--2022).}
         \label{fig:corpus-years}
         \Description[]{Frequency distribution of the number of complaints from our corpus by the year they were submitted starting from 2015 through 2022. The most complaints were from 2020 with 103 complaints, followed by 90 in 2019 and 83 in 2018.}
     \end{subfigure}
     \hfill
     \begin{subfigure}[t]{0.3\textwidth}
        \vskip 0pt
     \includegraphics[width=\textwidth, trim={7mm 2mm 21mm 1cm},clip]{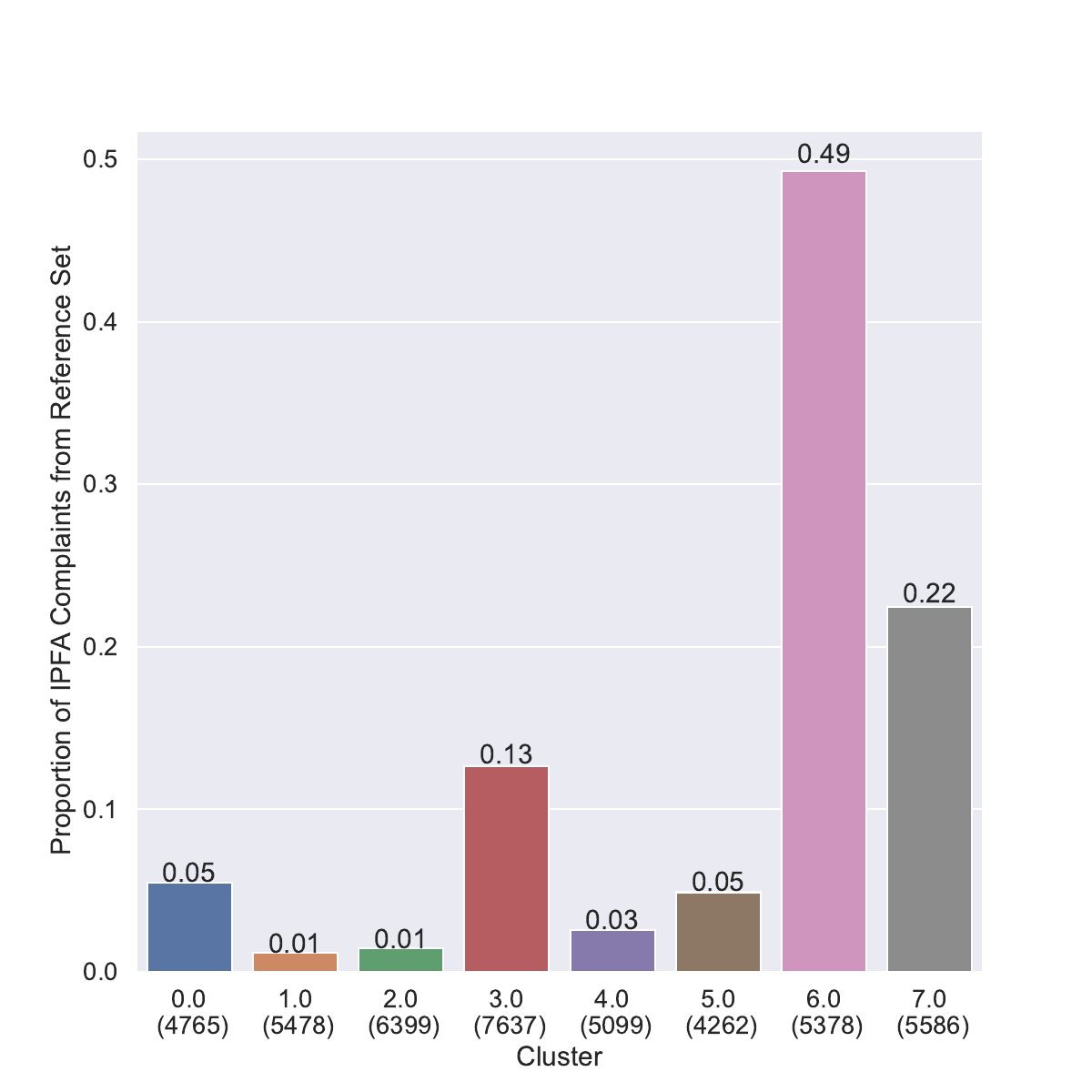}
    \caption{The distribution of \textit{ref} across clusters; where Clusters -3, -6, and -7 would be selected for review.}
    \label{fig:simcse}
    \Description[]{Distribution of reference set complaints in one iteration of our workflow. Cluster 6 has the highest proportion of reference set complaints with 49\%, followed by 22\% in cluster 7 and 13\% in cluster 3.}
    \end{subfigure}
\caption{The graphs from our meta-data analysis: the complaint length, distribution over years, and identification of higher \textit{ref} cases in clusters.}
\label{fig:three graphs}
\Description[Three graphs describing different aspects of our data corpus.]{}
\end{figure*}

Once embeddings of a sample of new complaint narratives and \textit{ref} were generated (\textbf{\circled{3}}), they were clustered using \textit{k}-means clustering \textbf{\circled{4}}, with the goal of grouping complaints that had semantically similar CCNs. \edit{We used the \texttt{sklearn} library's \textit{k}-means algorithm to perform this.}
Upon first attempt, we discovered that outlier-sensitive clustering algorithms, such as HDBScan \cite{Journalo11:online}, tended to exclude a large proportion of complaint embeddings from clusters due to the high dimensionality of the vectors and semantic complexity of complaints.
We thus relied on \textit{k}-means clustering, a process which assigns complaints to the same cluster if they are close (by distance metric) to a common cluster center \cite{schutze2008introduction}.
The clusters containing the most complaints from the current reference set \textit{ref} were then reviewed via their relative yield.
For example, in one iteration, we proceeded with the top cluster (\textit{Cluster-6}) which contained 49\% of our \textit{ref} complaints (\Cref{fig:simcse}), and cross-compared this with the second and third highest (22\% in \textit{Cluster-7} and 13\% in \textit{Cluster-3}).

\edit{These clusters were then sampled for complaints}, and were manually reviewed for IPFA relevance (\textbf{\circled{5}}) \edit{by five co-authors of this work, all with experience of IPV or at-risk populations in security contexts (see \textit{Ethical Considerations}). 
Each researcher received 300 sampled complaints per round across eight rounds (2,400 complaints/researcher total), and each complaint had two researchers for determining relevance.}
\edit{Complaints that no researcher marked as relevant were removed, and disputes were then reviewed by the group.} 
Determining relevance for reviewed complaints was challenging because of the need to understand IPFA's impact on the complainant and offender, as well to distinguish it from fraud and harassment (\Cref{sec:app-definitions}).

\edit{We augmented \textit{ref} with the reviewed complaints confirmed relevant and repeated the workflow (\circled{3}--\circled{5}).} After six iterations, this resulted in a final set of $n=\setdupe$ relevant complaints.

\paragraph{Workflow evaluation.}
Out of \setdupe IPFA complaints, 221 did not meet the keyword matching with proximity criteria (\textbf{\circled{1}}), and 169 did not satisfy keyword matching without proximity. This suggests that our workflow augmented the initial reference set, resulting in a 43\% increase in corpus size from \textbf{\circled{2}}. Our workflow did prove to require substantive time and effort on behalf of both expert and researcher reviewers, thus we look to supplement this method via other approaches to elevate this burden in future work (\secref{sec:discussion}).


During the workflow, SHapley Additive exPlanations (SHAP) scores --- a model feature explanation technique that reveals words or phrases that impact text classification --- helped identify words influencing a complaint's cluster assignment.
We calculated SHAP scores for \textit{Cluster-6} (mentioned in \circled{4}) using a random forest classifier trained on \textit{ref} \edit{(with the \texttt{sklearn} and \texttt{shap} libraries)}.
Term Frequency - Inverse Document Frequency (TF-IDF) vectors of each complaint were input to this classifier, with the complaint's cluster assignment as its classification.
Words like `our,' `we,' and `us' negatively affected a complaint's classification in the target cluster.
When complaints mentioned these words, it could indicate either that the complaint does not meet the definition of IPFA, or, it involves complex IPFA not easily identified due to shared financial harm.
The results suggest that our workflow can distinguish how complainants describe harm, which is crucial for identifying IPFA.

\edit{The \setdupe complaints produced with this workflow represent a sample of possible IPFA within the CFPB dataset. Given the reliance on keywords in creating our reference set (\textbf{\circled{1}}) and on our existing understanding of forms of financial abuse in manually reviewing clusters (\textbf{\circled{5}}), we emphasize that this dataset and methodology are not intended as a means of finding a sample of complaints that represent the entire space of how IPFA is described in complaints. Rather, it represents a focused dataset that can be further analyzed to inform future study exploring this larger space.}



\paragraph{Dataset and Framework Analysis.}
\noindent Following a manual review, a single author identified and removed a total of 49 complaints (2 duplicates, 47 re-submissions), associated with 37 unique identifiers.
This resulted in a dataset of \set narratives for further analysis.
Our complaints span from 2015 --- 2022, with the highest number submitted in 2020 (103), while the highest number of total complaints submitted to CFPB was in 2021 (\Cref{fig:corpus-years}) \cite{consumer_financial_protection_bureau_cfpb_2023}.
Most complaints in the corpus were short, containing 1K words or less (M:283.5, SD:333.8, \Cref{fig:corpus-lengths}), slightly longer than the average complaint length within the larger CFPB dataset (M:213.74, SD:259.06). Our shortest complaint contained a mere 15 words (``\textit{My former spouse unlawfully used my personal information to create an account with [institution]}'') (C327), and the longest contained 4,863 words.

In our analysis of our final dataset of \set complaints, we used a deductive approach of framework analysis \cite{becker_epistemological_2007} which is adaptable for specific questions \cite{furber_framework_2010}, limited timeframes \cite{kiernan_framework_2018}, pre-designed samples (e.g., customers of financial services) \cite{dixon-woods_using_2011}, and addressing a priori issues \cite{gale_using_2013}.
\edit{In framework analysis, data is sifted, charted, and sorted in accordance with key issues and themes through familiarization, identification of a thematic framework, indexing, charting, mapping, and interpretation.}
This method is optimal for datasets with a structured schema, like CFPB data categories, helping us summarize a complex dataset to answer our research questions.
\par To create an analytical \textit{``structure for guiding research''} \cite{eisenhart_conceptual_1991}, \edit{we used deductive reasoning to identify four essential components for indexing the narratives to differentiate an IPFA case from other financial harms.}
\edit{These components included: complainant identity, relationship status, financial product or service, and types of financial abuse reported by the complainant} (\Cref{tab:code-cat}, \Cref{sec:approach}).
\edit{For financial product or services, we obtained an initial list from the CFPB website that consists of a range of products that the consumer can identify in the metadata of the complaint.\footnote{The \href{https://www.consumerfinance.gov/data-research/consumer-complaints/search/}{Consumer Complaint Database} lists 74 sub-products.}}
\edit{To analyze the type of abuse, we relied on established} taxonomies on technology abuse \cite{freed2018stalker}, surveillance \cite{tseng2020tools}, technology-enabled financial abuse \cite{Bellini2023paying}, and economic abuse\cite{bruno_economic_2022} \edit{as a set of starting labels for developing our written descriptions.
By doing so, we were able to easily pinpoint instances of new types of technology-enabled abuse (discussed in \Cref{sec:analysis-attacks}).
Following a close reading of a subset of complaints ($n=100$), the research team regathered to determine how such abuse were discovered, the reported impact of such abuse, and what attempts (if any) that complainants made to resolve these concerns, were also poignant to capture.}

Five coders (all co-authors) independently assessed and \edit{indexed data from these eight categories} from each IPFA-related CCN in our dataset in three primary coding rounds over four weeks.
First, each coder \edit{coded} 20 cases each for consistency.
Inter-rater reliability was calculated, showing near-perfect agreement (Fleiss's $\kappa$: Product 0.91, Abuse 0.92, Discovery 0.52, Resolution 0.91) \cite{mchugh_interrater_2012}.
Discrepancies in \edit{how abuse was discovered (``Discovery'')} were resolved in round two (0.84). 
\edit{Due to such a high inter-rater reliability, the remaining 444 CCNs were coded by a single researcher}  followed by a final consistency check.
We then investigated how these variables intersected, by charting and interpreting any patterns or connections between categories via qualitative analysis software Atlas.ti to gain a clearer understanding and explanation of the `bigger picture' \cite{burgess_analyzing_1994}.\footnote{ATLAS.ti is a computer-assisted qualitative data analysis software that facilitates analysis of qualitative data for qualitative research, quantitative research, and mixed methods research,  \url{https://atlasti.com/}.}
\edit{Using a bespoke charting and summarization matrix (see \cite{furber_framework_2010, kiernan_framework_2018}), two authors identified higher level categories and typologies, which are presented in our findings (\Cref{sec:analysis-intro}).}

\paragraph{Critical Discourse Analysis.}
\edit{Understanding how survivors communicate about abuse is crucial for aiding survivors, as marginalized groups often avoid labeling their experiences \cite{ardener_ardeners_2005}. To answer our RQs, we had to dig deeper than merely reporting the descriptions complainants shared.
}
\par \edit{Critical Discourse Analysis focuses on how texts aim to persuade or convey meaning beyond the obvious to a human reader\cite{ritchie_qualitative_2013}, and we applied this approach to survivor's experiences of} financial abuse, its consequences, and areas for financial service provider intervention.
\edit{Thus, we applied Critical Discourse Analysis to scrutinize how complainants used language to convey what the financial abuse meant to them, and how they chose to communicate it.}
For instance, in our analysis, we would frequently read how complainants were distressed at feeling \textit{``unable to access finances''} or \textit{``having their finances controlled''} by an intimate partner, but did \edit{not report} such behaviors as explicitly abusive.

\edit{To do this, two authors performed} a close reading of each narrative, asking about intended meanings and significance.
\edit{This was done using the indexed framework analysis version of the dataset so relevant data could be easily located.
Two authors also made note of what elements were noticeably absent from the narrative and how complainants attempted to appeal to the emotional sensitivities of the reader.
Alongside identifying features, thematic elements and discursive fragments were also tagged and reviewed in a separate logging tool akin to analytical memoing.
}
\edit{To consolidate all these memos together, we used Atlas.ti for a second time to map a cohesive narrative across all these memos until we were satisfied we had captured all linguistic nuances of the narratives.}
Framework Analysis and Critical Discourse Analysis are complementary \cite{becker_epistemological_2007}, with Framework Analysis being epistemologically neutral \cite{burgess_analyzing_1994}, while Critical Discourse Analysis has been shown to surface underlying power relationships (a vital component of many IPV-related research projects) \cite{harris_im_2012, van_dijk_principles_1993}.

\begin{figure*}
    \centering
\begin{quote}
    \textit{I opened a \vio{credit card} with \tea{my bank} for rewards with an \tea{online retailer}. My \cya{ex-girlfriend} \pur{stole and used my card without} \pur{permission} and \ora{then owned up to using it once I found hidden charges} ... I initially agreed to let my \cya{ex-girlfriend} pay off the charges over time ... Later, I discovered that she had \pur{recorded my credit information and used it online without my knowledge} ... \maj{I contacted my credit card company to report the fraudulent charges}. Initially, they assured me that I was not liable for the charges and began the refund process. However, later on an investigator from \tea{the bank} reversed this decision, claiming that the \cya{ex-girlfriend} living in the same household \blu{made me responsible for the charges}. This resulted in a \blu{significant balance} on the \vio{credit card} and my \tea{credit company} reported a huge \blu{negative impact to my credit score} as a result ... \blu{I felt victimized} by both my \cya{ex-girlfriend}'s actions and the actions of \tea{my bank}.}
\end{quote}
    \caption{Example, paraphrased complaint (C5) with indicators for the framework creation. Our analytical framework is built from identifying the \protect\cya{abuser}, \protect\vio{financial product}, \protect\tea{actors}, \protect\pur{attacks}, \protect\ora{point of discovery}, \protect\maj{steps toward resolution}, and \protect\blu{negative impact} on the complaint.}
    \label{fig:example-complaint}
    \Description[An example complaint that has been paraphrased and underlined with indicators for our framework.]{Example, paraphrased complaint (C5) with indicators for the framework creation. Our analytical framework are built from identifying the abuser such as an "ex-girlfriend", financial product such as a "credit card" , actors such as "my bank", attacks such as "recorded my credit card information and used it online without my permission", point of discovery such as "I found hidden charges", steps toward resolution such as "I contacted my credit card company", and negative impact on the complainant such as "I felt victimized".}
\end{figure*}


\paragraph{Ethical considerations.}
As a \edit{cross} academia-industry team, we obtained approval from our Institutional Review Board and internal legal group before starting \edit{this work}.
Like other HCI scholars before us, we were cognizant that the use of public data incurs unique ethical concerns (e.g., de-anonymization, adversarial learning, and misrepresentation \cite{chancellor_taxonomy_2019, bellini:online}), particularly for the discussion of at-risk groups and sensitive topics \cite{warford_sok_2022}.
\edit{We thus} took steps to protect the digital safety of both the complainants and the research team.

The CFPB data we used is publicly accessible for research purposes, and may be downloaded without user registration on the CFPB website.
We chose to exclusively work with complaint data \edit{that was already redacted by the CFPB} \cite{consumer_financial_protection_bureau_cfpb_2023}. 
To ensure low risk of re-identification, each complaint was also manually checked for any uniquely identifiable details. 
\par We made a conscious effort to avoid any bias towards complaints from a particular time period, considering the entire history of the complaints database. 

We analyzed a one-year-old snapshot of the database (June 2022), which may allow complainants --- including from the newest narratives --- to have a greater chance to seek out safety resources. 
No effort was made to identify original posters.
We made no attempt to attribute complaints with their original identifiers, and have abridged prominent quotes to remove a few idiosyncratic details, phrases, or terms while retaining the meaning of the data to prevent reverse search-engine lookup.
Our approach involved human review and labeling by our research team, which built in discussions on distressing complaints during weekly meetings.
Each team member has extensive experience on researching the digital-safety concerns of at-risk populations. 
\edit{Three team members have a professional background in personally supporting and overseeing the voluntary service provision for survivors of IPV in security contexts.}
Thus, each researcher has a self-care strategy that mitigate the impacts of vicarious trauma \cite{moncur_role_2016}.

All work was conducted in secure, access-controlled cloud environments that were accessible to core research team members.
Finally, we also engaged two security professionals \edit{external to the research team} to review our work for the potential to teach adversaries new strategies or techniques to exacerbate their abuse. 
\edit{Both experts} judged that this work did not contain novel techniques viable for adversarial feedback.

\section{Findings}
\label{sec:analysis-intro}
\noindent 
In this section, we report the findings from two qualitative approaches (Framework Analysis, Critical Discourse Analysis), starting with an analysis of the profile of consumers who reach out to financial institutions with reports of technology-enabled financial abuse.
We then attempt to answer what attacks and devices are implicated \textbf{RQ2} (\Cref{sec:analysis-attacks}) and what barriers survivors encounter while doing so \textbf{RQ3} (\Cref{sec:analysis-barriers}).
\edit{Device and attack counts are provided for reference purposes, but should not be read as proportional attacks.}
Each example complaint has been lightly abridged and assigned an anonymous identifier (C1---C\set). \edit{We report on the proportion of descriptive codes used to characterize our dataset through percentages, as these codes were only used once per narrative and were mutually exclusive.}

\begin{table*}[t]
\small

\begin{tabular}{L{0.18\textwidth}L{0.29\textwidth}L{0.16\textwidth}L{0.28\textwidth}}
\toprule
\textbf{Relationship} & \textbf{Sample descriptors} & \textbf{Event} & \textbf{Sample descriptors} \\
\cmidrule(lr){1-2} \cmidrule(lr){3-4}
Ex-relationship, married & \textit{ex-husband (32.3\%), ex-wife (24.3\%), ex[-]spouse (7.1\%)} & Legal proceedings & divorce (45.7\%), theft (23.2\%), business disputes (9.4\%) \\
Ex-relationship, not married & \textit{ex[-]boyfriend (6.9\%), ex[-]girlfriend (4.3\%), ex (7.1\%)} & Parental responsibilities & childcare provision, alimony payments, visitation rights\\
Family members & \textit{father (3.7\%), mother (3.1\%), brother in-law (2.2\%)} & Criminal activity & identity theft (26.2\%), fraud (18.4\%), physical theft (15.2\%)\\
Other associates & \textit{close friends (3.4\%), ex-room-mate (2.2\%), new partner (1.3\%)} & & \\

\bottomrule
\end{tabular}
\caption{\edit{Most frequently occurring personal and contextual descriptors contained in our complaints. Where possible we have worked to demonstrate the proportionality of complaints that contain these, or lexical variations, of these descriptors. Note that ‘x gf’, ‘x girlfriend’, ‘x-girlfriend’.
all count as ‘ex-girlfriend’}}
\label{tab:relationships-events}
\Description[Most frequently occurring personal and contextual descriptors contained in complaints.]{Most frequently occurring personal and contextual descriptors contained in complaints. These are described by relationship types (i.e. Ex-relationship, married; Ex-relationship, not married; Family members; Other Associates), and Events (i.e. legal proceedings, parental responsibilities, and criminal activity).}
\end{table*}
\paragraph{\edit{Complainant and abuser profiles.}}
Our corpus gave us insight into the relationship between an abuser and complainant, their living situation, their roles in a family unit, and, the level of support the complainant was able to receive from financial institutions (\tabref{tab:relationships-events}).
\edit{A small number of complainants reported more than one abuser, individuals often who were able to exploit their close physical and emotional relationships with the complainant, including extended family members, close friends, and business associates.}
While we cannot claim that pronoun identification is an effective method to determine gender, we can highlight that, 
akin to other HCI work in this area \cite{freed2018stalker, bellini2021so}, complainants predominantly reported incidents involving a sole former male partner \cite{ali2021improving}, namely an ex-husband, and former non-married relationships, namely an ex-boyfriend.

Validating other identified risk factors as highlighted by HCI scholars \cite{freed2018stalker, tseng_digital_2021}, we identify a significant proportion of complainants reported either seeking legal representation or were already engaged in legal proceedings, primarily related to divorce.
This is perhaps to be expected, as complainants jointly described that divorce had been both the point of discovery for such abuse and a motivator to take action against such harms, often through hiring a legal advocate.
Just one complainant labeled their experiences as `financial abuse', validating our approach to alleviate the burden of relying on customers to self-report or self-identify financial abuse in their interactions with digital devices.

\paragraph{\edit{Complaint profile.}}
Technology-enabled financial attacks by an intimate partner was costly to a complainant, which resulted in substantial negative impacts on a complainant's financial well-being. 
We separate this harm into four distinctive categories: money loss from \textit{theft}, \textit{debt} incurred by identity theft, \textit{financial fees} (toil of IPV \cite{postmus2012understanding}), or \textit{finances that were withheld} from them by an abuser.
\par Thirteen complaints reported direct financial losses due to theft, totaling \$671,035 (M: \$51,618, SD: \$105,681), such as digital checks from a shared online business account (\$10,000) and unauthorized access to peer-to-peer payment accounts (\$6,000).
Additionally, 101 cases involved complainants being unaware of accounts opened in their name by abusers, resulting in a total reported debt of \$656,256 (M: \$6,497, SD: \$103,948), ranging from secured credit cards to substantial loans.
\par Technology-enabled financial abuse can also result in indirect costs which 10 complainants immediately had to pay to pursue or cover, such as legal fees, overdraft fees, and digital forensics totaling \$41,970 (M: \$4,197, SD: \$124,231).
Furthermore, complainants reported \$4,590,970 (M: \$124,000; SD: \$103,110) lost due to financial negligence, where abusers withheld money for services they were responsible for, such as child support and housing. \medskip

\subsection{What areas of digital financial products do complainants report abusers targeting?}
\label{sec:analysis-attacks}
\begin{figure*}[!htb]
    \centering
    \includegraphics[width=0.8\textwidth, trim={5mm 1cm 4mm 0},clip]{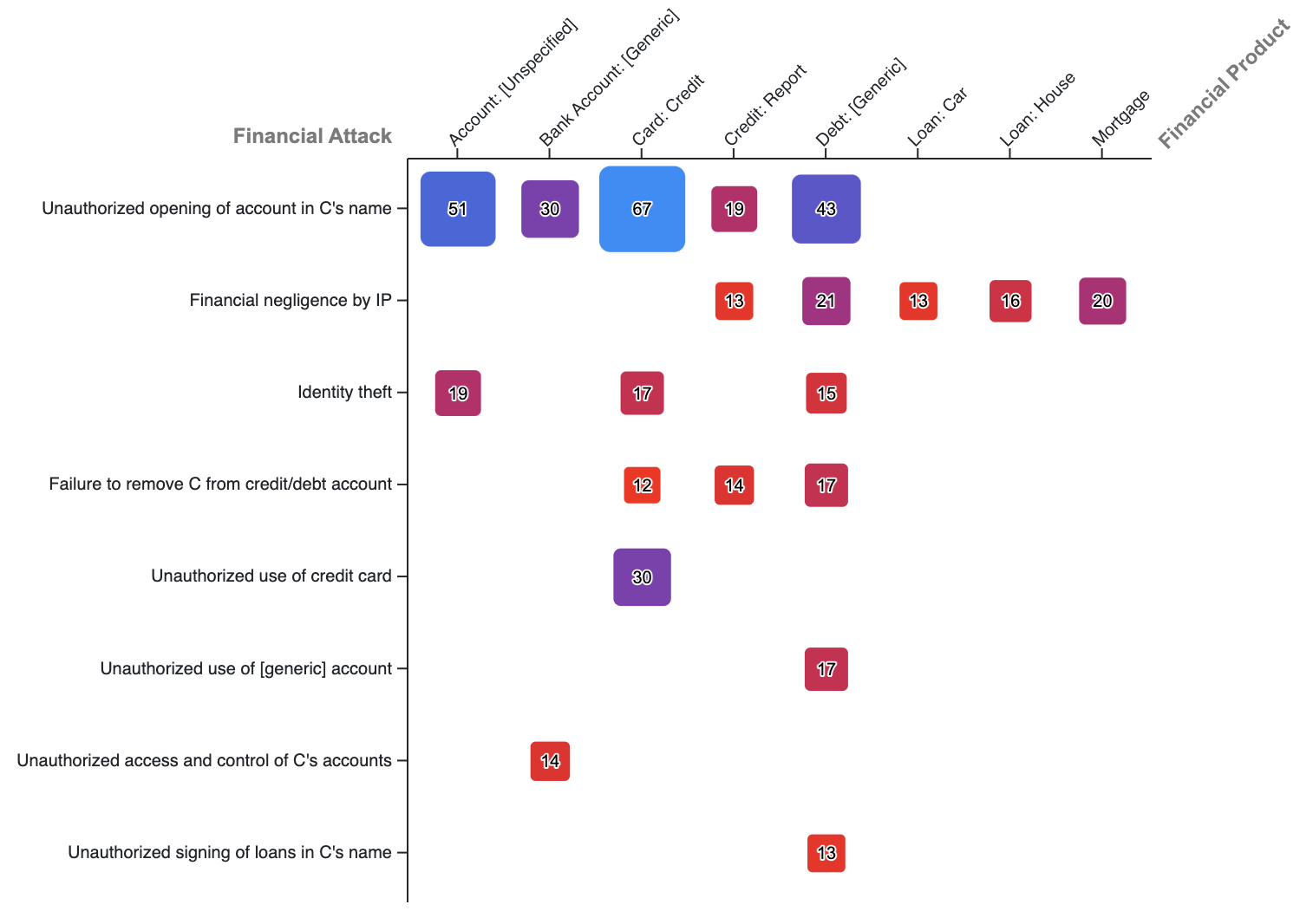}
    \caption{\small Top 20 products to financial attack mapping. Complainant (C), Intimate Partner (IP), [\textit{Unspecified}] means unspecified product, [\textit{Generic}] means a collective of consumer banking products.}
    \label{fig:attack-product-matrices}
    \Description[Diagram showing the mapping of the top 20 financial products to financial attacks from our analysis.]{Mapping the top 20 financial products from our analysis to financial attacks. The most frequent pairs were Unauthorized Opening of Accounts in C's name and Card:Credit with 67 occurrences. After that was Unauthorized Opening of Accounts in C's name and Account:Unspecified with 51 occurences. After that was Unauthorized Opening of Accounts in C's name and Debt:Generic with 43 occurences.}
\end{figure*}

\noindent Our analysis elicited 14 independent forms of technology-enabled financial attacks across 24 different technical products, systems, and services.
Among the \set complaints, the most common attack types were the opening of a checking or credit account in a complainants' name ($n$=186), negligence on financial duties to a complainant ($n$=141), and the theft of the complainants' identity ($n$=53).
Less frequently mentioned were an abuser conducting fraudulent chargebacks via a complainant's account ($n$=11), unauthorized request of their credit report ($n$=6), bankruptcy filing without a complainants' knowledge ($n$=6), and restricting a complainant from accessing their account ($n$=4). 
The attacks relied on named digital financial products (\Cref{fig:attack-product-matrices}) and roughly correlated with established technology abuse taxonomies in this space \cite{Bellini2023paying, freed2018stalker, tseng2020tools}.

We identify three areas of significance that had variations on known technical attacks (e.g., unauthorized opening of accounts \cite{freed2018stalker}, identity theft \cite{zou_examining_2020, zou_concern_2018}) that we suggest indicate novel manifestations of technology abuse. 
\edit{We present these by their overarching behaviors, such as negligence, account takeover, and deception.}
Namely, we identify that \edit{\textit{Access/account takeovers} for explicitly criminal activity}, \edit{\textit{Negligence} over asset/debt ownership}, and \edit{\textit{Deception and interference with customer-firm interactions}} were particularly devastating to complainants.

\paragraph{Asset/account takeover for criminal activity.}
Malicious blocking has been reported in prior studies to be an effective way of denying someone access to their own or shared accounts \cite{freed2018stalker}, which may often be performed through repeated password requests or through triggering an investigation into an account \cite{Bellini2023paying}. 
However, our analysis shows that abusers may leverage a hijacked account for other criminal purposes, namely, to conduct other acts of identity theft, scams, and more.
In these situations, an account takeover was not just performed to gain access and control over a complainant’s data or finances, but for abusers to leverage a complainant’s consumer products as a layer of protection against their own being implicated in fraud cases.
\edit{While some reports show that financial abusers use a survivor’s account that are in better financial credit standing than their own \cite{stylianou2018economic, Bellini2023paying}, we have yet to discover accounts in prior HCI literature such as these where abusers directly implicate them in a crime through this takeover:}

 \begin{quote}
     \textit{``I have discovered I am now a victim of identity theft ... My ex husband was deported for using my identity as well as my son's ... please help me resolve these online accounts on my credit report ... it is fraudulent and was not opened by me.''} (C269)
 \end{quote}

In some cases, this could mean engaging in reported money laundering, or using their complainant’s account as a means to commit acts such as charge backs, romance scams, or extortion.
\par Complainants in our dataset expressed disappointment upon finding out that their hijacked account affected multiple \edit{other targets of a crime}, especially since legitimate purchases could be falsely reported by the abuser.
Complainants described a range of worrying channels for discovering previously unknown forms of financial abuse, such as discovering `unknown accounts'.
\begin{quote}
    \textit{``... my online bank account was compromised by my ex husband and fraudulent activities were done against my account. I notified my bank ... when several thousands of forged money orders were deposited into ATMs ... I haven't committed a crime nor have authorities pursued any charges to justify why this ban for me to open a bank account ...''} (C28)
\end{quote}

Online sources of information on credit, such as a report (listing credit history), being contacted by a collections agency, and a credit score check history proved to be the most common discovery channels for such attacks, roughly co-aligning with how many complainants may discover instances of identity theft. Such findings suggest that despite prominent advice around identity theft \cite{zou_examining_2020}, some complainants do not discover financial abuse through traditional information discovery methods \edit{entailing that ``fraudulent activities'' could continue without their knowledge. 
Complainants that described} technology-enabled financial abuse, could, however already be aware of the existence of financial misconduct, or, were not prompted to reflect on how such information came to light (e.g., upon complaint submission).

\paragraph{Negligence over digital asset/debt management.}
Financial negligence in connection to digital products stands out from other forms of technology-enabled harm (e.g., \cite{tseng_digital_2021, freed2018stalker, Bellini2023paying} as it involves an abuser deliberately breaching a previously established financial agreement. 
These agreements may vary in nature, being either legally binding, such as child support or debt division in a separation, or informal, like the refusal to contribute to rent or utility bills.
Financial negligence could cover a wide range of actions, from refusing to negotiate an online agreement to breaking a digital asset arrangement.
In one instance, a complainant shared being unable to coordinate in the repayment of an outstanding debt due to being unable to contact their former partner:

\begin{quote}
    \textit{``I cannot submit any documents without my ex-husband of 14 years' signature ... This situation is extremely challenging. He harbors resentment, is uncooperative, and abusive.  He is aware that this issue affects both our credit reports, yet he adamantly refuses to assist in resolving it''} (C17)
\end{quote}

This could also extend to abusers reneging on a previously agreed-upon arrangement, most often through the court, which could be described by complainants as an avoidance to \textit{``adhere to stipulations''} (C59).
Many of these agreements, especially those stemming from legal separations, often include considerations of ownership related to specific assets or debts, such as being awarded property. 
The processes surrounding asset and debt ownership became a clear area in which this type of abuse intersects with financial institutions' services. 
For instance, we identified that complainants often suffer harm from this type of attack precisely because they are still held accountable when they believe they should not be.
When describing a challenge with respect to a shared online joint account, a complainant shared:
\begin{quote}
    \textit{``my ex-husband and I maintained a shared account together. According to our divorce agreement, he assumed responsibility for both the account and the associated debt. Despite this clear arrangement, my bank has been uncooperative in removing my status as a joint online account holder,  even though I am no longer legally liable for any debt accrued on that account since our date of separation ...''}
    (C222)
\end{quote}

{\small \begin{table*}[!htb]
    \centering
    \begin{tabular}{L{0.32\textwidth}L{0.33\textwidth}L{0.29\textwidth}}
        \toprule
        \textbf{Asset/Account Behavior} & \textbf{Asset/Debt Ownership and Management} & \textbf{Customer-Firm Interactions} \\
        \cmidrule(lr){1-1} \cmidrule(lr){2-2} \cmidrule(lr){3-3}
        \begin{itemize}[nosep, left=0pt,
                before={\begin{minipage}[t]{\hsize}},
                after ={\end{minipage}} ]
            \item Challenging valid transactions to create unexpected debt
            \item Deceptive handling of borrowed or stolen funds
            \item Intercepting funds for coercion or personal gain
            \item Unauthorized financial actions under the target's profile
            \item Unauthorized use of financial assets for fraud
        \end{itemize} &
        \begin{itemize}[nosep, left=0pt,
                before={\begin{minipage}[t]{\hsize}},
                after ={\end{minipage}} ]
            \item Blocking/reversing payments on financial agreements
            \item Failure to remove target from credit/debt account
            \item Ignoring/reverting established financial agreements
            \item Non-compliance with agreed financial obligations
            \item Refusal to establish financial agreements
            \item Tricking the target into signing financial obligations
        \end{itemize} &
        \begin{itemize}[nosep, left=0pt,
                before={\begin{minipage}[t]{\hsize}},
                after ={\end{minipage}} ]
            \item Exploiting financial hardship as leverage
            \item Threatening legal action, including divorce
            \item Using fear of other abuses as leverage
        \end{itemize} \\
        \bottomrule
    \end{tabular}
    \caption{Common attack types identified via complaints, organized by \edit{financial service it interacts with.} }
    \label{tab:taxonomy}
    \Description[Common attack types found in our analysis, divided by the relevant financial services they interact with.]{Common attack types found in our analysis, divided by the relevant financial services they interact with. The three financial services are: "Asset/Account Behavior", "Asset/Debt Ownership and Management", and "Customer-Firm Interactions".}
\end{table*}}
\paragraph{Deception and interference with customer-firm interactions.}
Our analysis uncovered several tactics employed by individuals engaging in financial abuse against their intimate partners that align with the concept of \textit{`financial deception'}, or \textit{`financial infidelity'} \cite{department_of_justice_elder_2023}.
However, in this context, we discovered it exhibits a distinctive, darker nature.
While many intimate partners might be driven by a desire to avoid upsetting their significant other or to evade confrontation \cite{karam_intimate_2023}, complainants explicitly characterize these actions as deliberate, explicitly harmful components of a broader pattern of abuse.
For instance, when bank statements were sent electronically, complainants reported uncovering instances where an abuser had set up email redirects to intercept emails from financial institutions, effectively preventing the statements from reaching them. 
In joint accounts, abusers occasionally prevented information sharing by providing their personal phone number for both accounts, ensuring that only an abuser  received notifications of any irregularities.

\begin{quote}
    \textit{``... I reached out to my bank's fraud department for assistance, but they declined to take any action on my behalf ... I neither initiated the creation of this credit card nor possessed any knowledge of its existence, let alone ever receiving a single statement. It came to light that my wife had confessed to concealing these online statements from me. ... My legal counsel advised me that, given my wife's admission to defrauding me and concealing this fact for a duration of three years, I should request the credit card statements ...''} (C361)
\end{quote}

In a few instances, complainants described instances in which abusers partially admitted to certain purchases, but complainants reported being misled about the actual cost of these purchases.
Another tactic involved abusers secretly maintaining a bank account unbeknownst to their partner, thereby circumventing the need to share funds equitably. However, in most cases outlined in our complaints, these hidden accounts were primarily used for incurring debt and engaging in high-risk borrowing, often on short notice. 
\par In extreme cases, we came across situations where a spouse had accumulated debts without the knowledge of the other spouse, and these debts only came to light following the debtor's death. 
In these scenarios, complainants expressed distress not only at having to cope with the loss of a spouse but also at having to address the joint debts in their name, debts of which they were previously unaware.

\subsection{What barriers to resolving financial abuse do complainants report encountering?}
\label{sec:analysis-barriers}
\begin{figure*}[!htb]
\scriptsize
     \centering
\vskip 0pt
\includegraphics[width=0.8\textwidth, trim={0 15mm 0 0},clip]{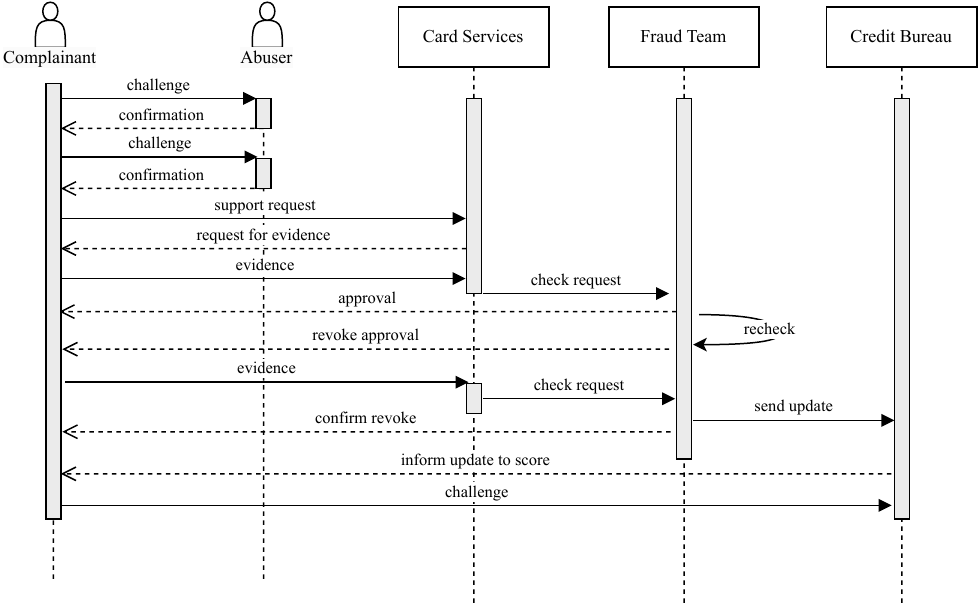}
\caption{Example complexity of multiple interactions between financial services inherent in a single case of IPFA. This model was generated via the account in \figref{fig:example-complaint}, where a complainant challenges their abuser twice about the use of their credit card details online, and coordinates between card services, fraud teams, and credit bureaus.}
\label{fig:complaint-journey}
\Description[Example of the complexity of multiple interactions between financial services and a complainant in a case of IPFA.]{Example complexity of multiple interactions between financial services inherent in a single case of IPFA. This model was generated via the account in Figure 4, where a complainant challenges their abuser twice about the use of their credit card details online, and coordinates between card services, fraud teams, and credit bureaus.}
\end{figure*}

\noindent A poignant reason for contacting the CFPB was that a complainant had received an unsatisfactory resolution from their financial institution, credit bureau, or consumer protection organization (17.9\%) --- a situation that CFPB's \textit{Customer Complaints Initiative} was explicitly designed for \cite{consumer_financial_protection_bureau_cfpb_2023, dehingia2022help}.
In the final phases of our analysis, we considered what \textit{types} of unsatisfactory resolutions (or lack of) may trigger a complainant to reach out to such services, as this has significant implications for how \edit{digital services are designed to improve this access.}
Our analysis shows that a single complaint journey has multiple stages, each necessitating substantive interaction with different financial entities (\figref{fig:complaint-journey}), with complainants often having to repeat the elements of the same story of technology-enabled financial abuse multiple times, or having a lack of access to important information regarding their finances.

As addressing barriers to socio-technical systems for recourse from abuse has significant implications, our analyses identifies three sub-types that constitute barriers to addressing technology-enabled financial abuse; first, an \textit{absent policy design to recognize financial abuse in intimate partnerships}, second, \textit{barriers to evidencing the existence of technology abuse}, and, most concernedly, \textit{a potential escalation of digital-safety risks through recommended resolutions by the financial institution}.
 
\paragraph{Policies that overlook the dynamics of technology-enabled financial abuse.}
\edit{Mirroring work in other areas of digital finances in HCI \cite{cunningham_cost_2022, surviving_economic_abuse_conversation_2021}}, we identified that many complainants reported feeling that they were poorly served by company policy in managing financial abuse cases. 
We identify two commonly occurring challenges in the context of IPV: authorized transactions and consent-based concerns.
\par Identifying the difference between authorized and unauthorized online transactions was a common problem for many complainants; where receiving a judgement of unauthorized would trigger a fraud investigation, while authorized could result in a lack of action.

Although peer to peer payment services were a rare occurrence in our dataset (as many of those companies do not as of yet earn over \$10B in assets\edit{, the minimum threshold to be tracked by the CFPB}), we identified several complainants who cited previous legitimate interactions with an abuser \edit{where the financial service provider refused to label} fraudulent transactions as fraudulent due to a previous history between the users. We noted a lack of consistency in how a complainant's experience was categorized in our dataset.
For instance, one complainant shared the following after having a phone stolen:

\begin{quote}
    \textit{``... there was a fraudulent transaction from my account issued to ``[fake-name]'' in the amount of {\$10,000.00}. This ``[fake name]'' was an ex-boyfriend who had stolen my phone and sent the money to himself after turning off my notifications to my number. I immediately contacted [P2PP] and inform them of the fraudulent transaction ... I was able to get nothing resolved until far after the time to which they informed me that it was a non fraudulent transaction because said `` [fake name] '' and I had previous transactions''} (C97)
\end{quote}
In this situation, the fact that a complainant had previously interacted with an intimate partner in this context --- despite a clear description of device compromise --- was enough to categorize this transaction as non-fraudulent; resulting in a substantive loss of money for a complainant.
\par In some cases, while complainants stated that employees of financial institutions were more sympathetic to their concerns, most reported an inability for policy to permit them to do so:

\begin{quote}
    \textit{``My husband has 100\% control of all of our money ... I attempted repeatedly to get into my account to get funds to hire an attorney ... I was then told I never had an account there ... I pleaded with the teller who quietly told me she believes me but she would lose her job if she tried to help me''} (C251)
\end{quote}

\paragraph{Challenges to evidencing `legitimate' financial abuse.}
Despite multiple adversities, some complainants attempted to take on the complexities of their specific financial case independently, but were met with a \textit{lack of a response} from the financial organization (7.3\%).
We identified that the threshold for being able to demonstrate proof of financial abuse was considerably high, namely, that when complainants attempted to contest authorized opening of credit cards in their name, a financial organization would turn around and ask for `proof':

\begin{quote}
    {\textit{``... My ex-wife has continued to obtain new credit in my name. I have disputed many accounts that appear on all 3 credit bureaus only to have them respond with ''we need proof''. ... nothing changes or helps ... I'm really just stuck ... This is insane.''}} (C48)
\end{quote}

The quality of the evidence was also judged harshly; oftentimes, screenshots and transcripts were rarely taken into consideration, as they were judged to not meet the threshold expected to dispute such harms.
Complainants sadly provided detailed descriptions of trauma as a consequence of collecting their own evidence  (2.4\%).
Digital evidence gathering was distressing for many complainants who reported that they had already been affected by similiar requests during legal proceedings, such as for divorce or restraining orders.
However, this reportedly trapped them in what a complainant described as a ``vicious cycle'' --- that to get the digital proof they needed to ask the financial institution to send over the documentation, whom would then refuse to do so without first seeing `proof' of abuse.

While some financial institutions aimed to assist complainants with the challenge in documenting their experiences, this also introduced entirely new barriers.
Many complainants reported not receiving necessary documents or lacking the means to obtain them from financial institutions, or feeling pressured to share personal information they would rather keep private (e.g., salary, location history).
In one scenario, a complainant received the requested statements relating to an abuser opening an account via an email from a financial institution, but encountered a further digital barrier, reporting they were \textit{``unable to open the zipped attachment to see the contents''} (C89) as they required a password that the complainant reported the institution then refused to provide.

\paragraph{Suggested resolutions may elevate digital-safety risks.}
Many financial products we identified in our analysis were often co-shared or connected between a complainant and an offender, including bank accounts, house loans, car insurance, and utility bills.
Relationships with offenders can present unique digital-safety risks \cite{warford_sok_2022}, such as fraudulent purchases, identity theft, or social engineering to elicit financial information through social engineering \cite{wood_financial_2017, freed_digital_2017, freed2018stalker}.
Our analysis revealed that \edit{financial support workers} are broadly unaware of such a risk, and made multiple recommendations for a complainant to directly \textit{challenge} an offender until the concern was resolved.
In spite of explaining concerns about being unable to resolve the issue on their own, \edit{financial support workers} and branch staff would recommend complainants \textit{``take the issue up''} with an offender (C11).
As one complainant described after reporting a card opened in their name:

\begin{quote}
    \textit{``I called today to find out the status, and was told [by the \edit{financial support worker}] that I benefited from this. I don't understand how I benefited from being the victim of domestic abuse. They said I had to file a police report, but, if I did that, my husband could possibly hurt me physically ... ''} (C82)
\end{quote}

In general, complainants reported being afraid of physical retaliation after approaching the abuser, citing \textit{``feeling fearful and scared''} of the person that they had been admonished for \textit{``trusting them''} by a financial representative (C306).
Namely, complainants were often encouraged by representatives at financial institutions to find another source of income, forgo claims for theft of financial assets, hire an attorney to pursue legal action, to \textit{``find another bank''} (C408), or even to pay off the outstanding debt.
In response of being recommended to modify a loan took out through coercion (`coerced debt') by a representative at a financial institution, one complainant exclaimed: \textit{``I am not looking to modify this loan, I am looking to be free of this loan!''} (C288).
\par While \edit{financial support workers} should not be expected to provide direct IPV support to complainants who describe this (as highlighted by Zou et al. \cite{zou_role_2021}), such results point to the need for \edit{financial support worker}s to have training on how to handle such complaints without discrediting or putting a complainant at further risk.

\section{Discussion}
\label{sec:discussion}
\noindent \edit{Through a combination of natural language processing techniques and careful human review, we have taken the first step in what identifying consumer complaints linked to IPFA could look like in public datasets (\textbf{RQ1}).}
\edit{Our subsequent analyses provides a complementary approach to existing work in HCI on technology abuse \cite{freed_digital_2017, freed2018stalker} by uncovering new insights into digital products (\textbf{RQ2}), and barriers to digital services (\textbf{RQ3}), from survivors who do not label their experiences as abuse, and thus, may not presently be receiving professional help.}

\edit{Despite the challenges of complicated reporting pathways (\Cref{sec:analysis-barriers}) \edit{and navigating the design of} digital products that do not accommodate their digital-safety needs (\Cref{sec:analysis-attacks}), survivors are nevertheless still reaching out to customer complaint services for help.}
\edit{In light of these findings, we recommend several strategies to bolster the digital-safety of individual financial products through improved security tooling (\Cref{sec:discussion-checks}), enhance digital systems for evidence gathering (\Cref{sec:discussion-logs}), and suggest ways to improve financial support systems to support clients (\Cref{sec:discussion-fsw}).}

\subsection{Safety Checkups for Digital Financial Accounts}
\label{sec:discussion-checks}
\noindent Our research reveals that unclear account ownership and authorization approaches can create unique vulnerabilities to survivors of IPFA attacks in digital financial systems.
These are common, complex challenges for any digital system \cite{levy_privacy_2020}, but the negative consequences of getting this wrong for survivors are severe.
\edit{The unauthorized use of a complainant's accounts for criminal activity resulted in the complainant being held accountable for the repercussions  (\Cref{sec:analysis-attacks}).}

Similarly, when it comes to financial deception, the complainant is unfairly liable for a joint account they had no knowledge of.
While prior HCI research has already highlighted the complications that may arise from close social relationships --- be these unofficial proxies \cite{latulipe_unofficial_2022} (adults who assist older adults) or caretakers \cite{mentis_upside_2019} --- intimate partners may have to pool their financial resources to pay for specific assets or experiences \cite{mcdonald_realizing_2020, Bellini2023paying}, which require shared account access and permissions \cite{park_can_2020, HowcanIs54:online}.
\edit{Revoking consent for shared access is not effectively indicated in most digital account or asset workflows, or notifications to users, making it difficult to identify attacks until much later.
Thus, focusing on improved digital workflows could benefit the user's understanding of account ownership and where their information is being used elsewhere.}

\textit{Safety checkups} geared towards financial systems may be able to help survivors or other concerned users audit their mobile and desktop devices for signs of technology abuse. 
Such safety checkups are initiated by users and help them with actions such as checking and resetting app permissions, configuring privacy settings, and monitoring device logins \cite{daffalla_account_nodate}. The user interfaces for carrying out the above actions exist across different technology platforms \cite{daffalla_account_nodate}, and was even observed by authors in existing financial mobile apps. However, a guided workflow for utilizing these interfaces is not always present. 
\par A prominent example of such a guided workflow is Apple's Safety Check feature for the iPhone \cite{apple_safety_check}, which guides users through a review of: data (e.g. location, shared photo albums) users share with other users, app permissions, and devices logged into a user's Apple ID. This feature also provides users the option to reset permissions and sharing privileges, providing users with a description of the potential consequences of doing so (e.g. other users may be able to observe a loss of access).
Based on our analysis, it is evident that digital financial workflows may benefit from similar guided mechanisms being implemented in financial institution applications that build on the actions that existing safety checks already employ. 

Regularly auditing financial accounts with such a guided tool could help identify early signs of financial attacks, such as by looking for signs of unexpected account behavior or how personal financial data, credit cards, and other assets have been attributed elsewhere without permission. For instance, a financial safety check mechanism may expand on reviewing device logins by exposing more details about account accesses during in-branch or over the phone interactions. This could potentially reveal acts of financial deception where an offender attempts to impersonate the target while speaking to a relevant financial service provider.
Further, if any suspicious activity is identified, users could receive trauma-informed recommendations on their next actions \cite{zou_role_2021}, like notifying their financial institution or initiating the resolution process.

Financial institutions are already well-equipped to provide such support to other groups vulnerable to digital financial crimes, such as targets of identity theft or scams.
Thus, it could be the case of adapting existing mechanisms to fit this specific use case, particularly during high-risk events such as during account openings or when fraud policies are triggered. We see particular promise in situations involving \textit{deception and interference with customer-firm interactions} (\Cref{sec:analysis-attacks}). Thus, we are confident such functionality would also benefit other, inter-related financial harms, such as account compromise through device theft, or elder financial abuse ~\cite{latulipe_unofficial_2022}.

\subsection{Automated, intelligent approaches to technology-enabled financial abuse evidence}
\label{sec:discussion-logs}
\noindent \edit{Insufficient clarity on what constitutes sufficient evidence and complicated resolution procedures can obstruct the evidence gathering and resolution process for possible IPFA cases (\Cref{sec:analysis-barriers}).
Thus, digital evidence that validates a survivor’s experience of abuse is crucial to address psychological harm \cite{postmus_building_2023, sharp2015review} and pursue a potential resolution \cite{freed_digital_2017, bellini2023digitalsafety}.
These challenges identified in our analysis, if left under-addressed, could directly increase consumer vulnerability, resulting in long-term financial costs in the hundreds of thousands (\secref{sec:analysis-intro}).
As digital evidence gathering processes may be required by financial institutions or legal policies to take any step towards resolving actions, we pose two approaches that can augment existing evidence gathering and evidence reporting approaches.}

\paragraph{Evidence gathering approaches.}
\noindent \edit{The process for gathering digital evidence for complainants can be difficult due to the dynamic nature of financial attacks. 
For instance, complainants reported losing digital documentation, unresponsive organizations, and cases where financial institutions provided the information in a format that was inaccessible (\secref{sec:analysis-barriers}).
These processes can make resolution to such attacks seem impossible due to complainants feeling overloaded by working with different stakeholders and unclear standards for sufficient evidence \cite{freed_digital_2017}.}

To mitigate these barriers to access, our results suggest that having a shared evidence log that adequately documents what is required to showcase different types of technical attacks, and the required standard that such digital evidence would need to meet could be useful.
\edit{While similar concepts already exist in cases of identity theft \cite{moskovitch2009identity}, the majority of digital financial harms that are largely exempt from our work, such as fraud and scams, are usually addressed as individual incidents.}
Our approach underlines that this shared evidence store should emphasize the complex social dynamics of intimate partner violence and the associated risks.
\par For instance, Surviving Economic Abuse, a registered charity that supports women who have experienced economic abuse by a current or former partner, offers \textit{The Economic Abuse Evidence Form} \cite{surviving_economic_abuse_economic_nodate} as a tool for debt advisers to consolidate information about abuse in a single location about the abuse experienced by a survivor, and help a debt adviser support a survivor when communicating with creditors.
While this digital form is designed solely with coerced debt in mind, we posit this could be augmented to cover more forms of technology enabled financial abuse, such as the types of financial attacks (see \tabref{tab:taxonomy}).
Such an approach could also inspire new frameworks for conversations about financial abuse and training for staff, as hinted at by Bellini \cite{Bellini2023paying} to simultaneously build confidence in responding to reported cases.

\paragraph{Evidence reporting approaches.}
The emergence of technology-enabled IPFA brings many new challenges to HCI scholars who are interested in alleviating the harm caused by financial and monetary loss, such as through preventing access to technical products (\secref{sec:analysis-attacks}), or the accumulation of fees or credit (\secref{sec:analysis-intro}). An obstacle that caused significant time burdens for complainants was that even when they had successfully collected evidence of suspected financial abuse, the reporting infrastructures introduced new barriers to sharing this with financial institutions, evidenced often by a lack of response by the institution to these efforts.
\par While there are many ways to approach the challenge of reporting abuse, a small design decision that could have an enormous impact on survivors \cite{strohmayer_trust_2021}, could be a closer look at the design of online complaint forms.
Doing so could help to gather further insight into the fine-grained information that were regrettably absent from our results.
The complaint form for the CFPB had a welcome large character limit for complainants to share experiences (10,000 words).
Though improved complaint form design that asks for vital contextual information --- potentially guided by the eight categories in our framework analysis --- could elicit good indicators for when evidence may be absent, it could also stall a process of resolution. 
\par Further, the form could ask optional questions about the complainant's relationship to the suspected offender, the impact of abuse, and their desired resolution, which could be encouraged by a natural language processing approach, akin to smart email suggestions, that could help the consumer identify what vital contextual information or digital documentation may be missing from their stories. Guiding complainants to provide details such as how they discovered such abuse may also be a valuable area of future study, with implications in identifying the appropriate means of support for a complainant, based on their unique circumstances.
While this technical suggestion is especially useful in the context of IPFA, this could also be useful more broadly to any complainant reaching out about financial challenges.

\subsection{Further insight into technology-enabled financial abuse}
\label{sec:discussion-fsw}
\noindent Existing computer security support infrastructure has been shown to fall short in cases of addressing the unique threat model of intimate partner technology abuse \cite{freed2018stalker, freed2019my, daffalla_account_nodate}, resulting in the development of carefully designed security clinics \cite{tseng2022care, havron_clinical_2019}.
\edit{However, as raised by Zou et al. \cite{zou_role_2021}, sadly many survivors may never reach these clinics. 
Our work provides a first look into how survivors of IPFA describe their own experiences independently of professional services, while also not identifying as a survivor of abuse.}
We are encouraged by the many suggestions offered by the HCI community \cite{bellini2023digitalsafety, Bellini2023paying, zou_role_2021}, that call for adequate training of financial support workers and frontline staff around technology abuse.
As our findings show, the fact that complainants reach out to multiple institutions over time (\figref{fig:complaint-journey}) while experiencing multiple attacks \cite{freed2018stalker, tseng2022care}, only reinforces this need.
\edit{However, we believe that further insight is still needed into IPFA to ensure that such interventions are effective.}

Our dataset is the first example of a corpus demonstrating potential IPFA cases in digital financial products `in the wild'.
\edit{Despite this, we acknowledge shortcomings of this dataset, and encourage the augmentation of this resource by data from other consumer organizations and financial institutions alike.}
When studying or addressing IPFA, we suggest that a broad range of complainant language should be anticipated and a focus on the attributes of the complainant experience may prove to be key in identifying supportive services, whether that be through manual or automated methods.
\edit{Specifically, merging our dataset with internal complaints from financial institutions can achieve a similar goal as monitoring tech-enabled IPV cases for computer security customer agents: capturing a detailed record of encountered attacks \cite{tseng_digital_2021, Bellini2023paying}, vulnerabilities in support systems \cite{bellini2023digitalsafety}, and the impact of trauma on targets and financial support workers~\cite{zou_role_2021}.}

Building and analyzing such a dataset at scale may benefit from triaging large complaints datasets at financial institutions, possibly with automated approaches.
For instance, one could leverage complaints language and recent techniques that demonstrate performance gains in language modeling tasks with large language models, step-by-step reasoning, and rationales \cite{hsieh2023distilling, zhang2023multimodal, huang2022large} to create a robust classifier of complaints.
\edit{A cohesive set of IPFA examples across different consumer reporting sources can help researchers as well. 
Research characterizing the harms of IPFA over time may be improved by joining a survivor's complaints made to different organizations as abuse progresses.
Similarly, non-profit organizations for survivor assistance and advocacy may better support survivors in their evidence gathering and reporting approaches with a unified evidence store.
These use cases all call for thorough investigation prior to implementation, but we hope that they inspire stakeholders in HCI to enhance system safety for all.}

\paragraph{Limitations.}
\noindent Our dataset is modest compared to some fraud datasets \cite{ftc_data_2013, ftc_explore_2019}, so our results should be interpreted cautiously by designers, developers, and fellow researchers.
\edit{Namely, our study focus may not be representative of all survivors, as our dataset only examines a specific sample of IPFA cases where the complainant explicitly shows awareness of harm or abusive behavior through specific keywords.}
The subsequent use of our workflow intends to mitigate this issue, but still relies on the results of \edit{proximity-based keyword-based approaches as a reference set} which may limit how new patterns of IPFA are discovered.
Like other HCI studies based on self-reported data, complainants may overstate certain aspects of their financial history due to social desirability bias \cite{bhatia2022complaint} or to elicit empathy and alleviate financial burdens \cite{gambetta2015complaints}.
People with negative experiences may be more inclined to write formal complaints \cite{joudaki2015using}, while others may fear repercussions or societal stigma~\cite{freed_digital_2017, sharp2015review}.

Our primary goal was nevertheless to understand what specific financial attacks and barriers complainants report experiencing to professional organizations.
\edit{Thus, despite these limitations,} our dataset still offers a substantial cross-section of survivor accounts to U.S.-based financial institutions, providing valuable insights for institutions dealing with financial abuse concerns, particularly via digital technologies.
We look forward to further work that explores the different forms of harm and language used in self-reported instances of IPFA.

\begin{acks}
\noindent We thank Fannie Liu and Francesca Mosca for their feedback on this work. We are also grateful to our associate chairs and reviewers, whose comments helped to improve the manuscript. 

Rosanna Bellini's contributions are supported in part by NSF Award CNS-1916096 and a research award from JPMorgan Chase.


This paper was prepared for informational purposes in part by the CDAO group of JPMorgan Chase \& Co. and its affiliates (``J.P. Morgan'') and is not a product of the Research Department of J.P. Morgan. J.P. Morgan makes no representation and warranty whatsoever and disclaims all liability, for the completeness, accuracy or reliability of the information contained herein.  This document is not intended as investment research or investment advice, or a recommendation, offer or solicitation for the purchase or sale of any security, financial instrument, financial product or service, or to be used in any way for evaluating the merits of participating in any transaction, and shall not constitute a solicitation under any jurisdiction or to any person, if such solicitation under such jurisdiction or to such person would be unlawful.  
\end{acks}

\bibliographystyle{ACM-Reference-Format}
\bibliography{proceedings,sample}

\clearpage
\appendix

\twocolumn[
    \section{Determining intimate partner financial abuse}
    \label{sec:app-definitions}
    \vspace{1.5em}
    \noindent\begin{minipage}{\textwidth}
        \centering
    {\scriptsize
    \begin{tabular}{L{0.15\textwidth}L{0.18\textwidth}L{0.18\textwidth}L{0.18\textwidth}L{0.18\textwidth}}
    \toprule
         & \textbf{Intimate Partner Financial Abuse} & \textbf{Elder Financial Abuse} & \textbf{Financial Fraud} & \textbf{Financial Harassment}\\
    \midrule
         \textit{Adversary} & An individual \cite{Bellini2023paying} & An individual \cite{storey_risk_2020} & An individual, group, or business entity \cite{soltes2017fraud} & A set of individuals (organization) \cite{stace2021debt} \\
    \cmidrule{2-5}
         \textit{Adversary goals} & To financially benefit, to have power over the target, damage their reputation, or cause them harm \cite{postmus_building_2023, Bellini2023paying} & To financially benefit \cite{department_of_justice_elder_2023} 
         & To gain financial advantage often via deception \cite{consumer_financial_protection_bureau_cfpb_2023} & To 
         financially benefit often via intimidation \cite{peterson2015consumer} \\
    \cmidrule{2-5}
         \textit{Pre-Existing Trust Relationship} & Yes & Yes & No & No \\
    \cmidrule{2-5}
         \textit{Examples} & A partner opens non-consensual financial accounts in a target's name & An adult child uses a target's credit cards without their permission
         & A stranger promises high returns for a target's investment, but instead steals it & A collection agency harasses a debt owners' families to collect debts \\
    \bottomrule
    \end{tabular}
        \Description[Matrix characterizing Intimate Partner Financial Abuse, Elder Financial Abuse, Financial Fraud, and Financial Harassment]{Comparative matrix that characterizes each form of financial harm by the adversary involved, their goals, and the pre-existing trust between the adversary and target. The bottom row briefly describes examples of how financial assets and/or information are used in each form of harm. Intimate Partner Financial Abuse requires pre-existing trust relationship with an Intimate Partner, and the adversary may have goals involving harming the target, causing reputational damage, financially benefitting, or exerting control.}}
        \captionof{table}{Comparative matrix that characterizes each form of financial harm by the adversary involved, their goals, and the pre-existing trust between the adversary and target. The bottom row briefly describes examples of how financial assets and/or information are used in each form of harm.}
    \label{tab:dependencies_new}
    \end{minipage}

]


As many different groups can experience financial harm, we first identify the specific socio-technical characteristics that differentiate financial abuse in IPV context from other predominant harms. 
IPV has specific socio-technical characteristics that differentiate it from other forms of financial harm, including fraud, harassment, and the abuse of older adults (elder financial abuse) \cite{deliema2018elder, mentis_upside_2019}.
To begin with, IPV is characterized by a range of complex social behaviors that occur within intimate relationships, often by leveraging fine-grained details about the partner which may not occur across all forms of financial harms.
Following an in-depth literature review, we delineate these via adversary type, adversarial goals, and the existence of trust in a relationship (\tabref{tab:dependencies_new}).



\textit{Financial fraud} is committed by an individual, or group of individuals (e.g., an organization) to extract funds from a target with whom they do not have a pre-existing interpersonal, trust relationship \cite{joudaki2015using,smith2000fraud, anderson2004consumer, soltes2017fraud}. 
Deception is often used to achieve the adversary's main goal, which is financial gain, such as selling a product under false pretenses (e.g., undisclosed interest rates).
Conversely, \textit{financial harassment} involves a group of individuals (organization) with whom the target does not have a pre-existing interpersonal, trusting relationship. Offender objectives are primarily financial, and deception may  be used to achieve them, such as a debt collector making unsolicited calls to a debtor or their family and making unsubstantiated claims about the consequences of not paying the debt \cite{stace2021debt}.
\textit{Financial abuse} necessarily involves a single individual, the abuser, with whom the target has a pre-existing interpersonal, trust relationship. The abuser may seek to harm a target by exploiting, sabotaging, restricting, or monitoring a target's activity related to money.
Two prevalent sub-types of this form of harm are the financial abuse of elders, and financial abuse of intimate partners.
\textit{Elder financial abuse} is a form of elder abuse targeting the financial assets of vulnerable elderly adults (often defined as aged 60 or higher), and is recognized at a federal level \cite{department_of_justice_elder_2023}.
Elder financial abuse (EFA) is typically perpetrated by individuals with a pre-existing trust relationship with an elderly target (such as a caretaker-patient relationship friend, or family member \cite{darzins2010financial, deliema2018elder, nguyen2021perceived,noauthor_metlife_nodate}). 
Elder financial abuse may not rely on deception, and often takes advantage of \textit{cognitive decline} with age \cite{deane_elder_nodate}.
As cognitive decline and the accumulation of wealth and assets are both positively correlated with age \cite{spreng2016cognitive, han2016mild, deane_elder_nodate}, older adults are at an acute risk of being targetted by financially-motivated adversaries.
Finally, \textit{intimate partner financial abuse} (IPFA) occurs in the context of a romantic partnership. 
An offender may have goals beyond financially benefiting themselves, such as a set of complex social aims around the misuse of power and control over a target \cite{freed_digital_2017, Bellini2023paying}.
While IPFA can co-occur with EFA \cite{storey_risk_2020}, we differentiate between these forms of abuse based on the adversary involved and the adversary's  motivation 
to \textit{harm} the target's financial health --- a factor that would be counter-intuitive to the financial exploitation inherent to EFA \cite{Bellini2023paying}.

\section{Query Design and Keyword Search}
\label{sec:app-query}

\begin{table}[!htb]
    \centering
    {\small \begin{tabular}{L{0.9\columnwidth}}
    \toprule
    {\tt intimate partner keywords}\\
    \midrule
    ``spouse'', ``ex-spouse'', ``husband'', ``wife'', ``ex-husband'', ``ex-wife'', ``other half'', ``girlfriend'', ``boyfriend'', ``partner'', ``ex-boyfriend'', ``ex-girlfriend'', ``ex-partner``, ``fiance``\\
    \bottomrule
    \end{tabular}}

    \caption{Intimate partner keywords (\textit{intimate partner keywords}) we used in our study.}
    \label{tab:ipkeywords}
    \Description[]{Intimate Partner keywords (ipkw) we used in this study such as "spouse", "ex husband", "ex girlfriend", "other half".}
\end{table}
\begin{table}[!htb]
    \centering
    {\small \begin{tabular}{L{0.9\columnwidth}}
    \toprule
    {\tt financial abuse keywords}\\
    \midrule
    ``Steal'', ``Stealing'', ``Stole'', ``Stolen'', ``Hid'', ``Hide'', ``Hidden'', ``Spy'', ``Spied'', ``Spying'', ``Surveil'', ``Surveilling'', ``Surveilled'', ``Control'', ``Controlled'', ``Controlling'', ``Harass'', ``Harassed'', ``Harassing'', ``Abuse'', ``Abusive'', ``Abusing'', ``Abused'', ``Exploit'', ``Exploitative'', ``Exploiting'', ``Exploited'', ``Harm'', ``Harmful'', ``Harmed'', ``Harming'', ``Hurt'', ``Hurting'', ``Upset'', ``Upsetting'',  "Sabotage'', ``Sabotaged'', ``Sabotaging'', ``domestic abuse'', ``fraudulent'', ``fraudulently'', ``abused'', ``abusive'', ``violence'', ``violent'', ``stole'', ``stolen'', ``stealing'', ``forced'', ``harassed'', ``unwanted'', ``coerced'', ``opened'', ``victim'', ``victims'', ``survivor'', ``survivors'', ``Batterer'', ``Batterers'', ``perpetrator'', ``perpetrators'', ``abuser'', ``abusers'', ``Batterer'', ``Batterers'', ``perpetrator'', ``perpetrators'', ``abuser'', ``abusers'\\
    \bottomrule
    \end{tabular}}
    \caption{A full list of financial abuse key words (\textit{financial abuse keywords}) we used in our study.}
    \label{tab:abusekeywords}
    \Description[A full list of financial abuse key words fnkw we used in our study such as "stole", "abused", "exploit", and "harass".]{A full list of financial abuse key words fnkw we used in our study such as "stole", "abused", "exploit", and "harass".}
\end{table}

\newpage

\section{CFPB Schema and Workflow}
\label{sec:app-cfpb-workflow}

\begin{figure}[!htb]
\scriptsize
     \centering
\begin{subfigure}[b]{0.23\textwidth}
    \centering
    \includegraphics[width=\textwidth]{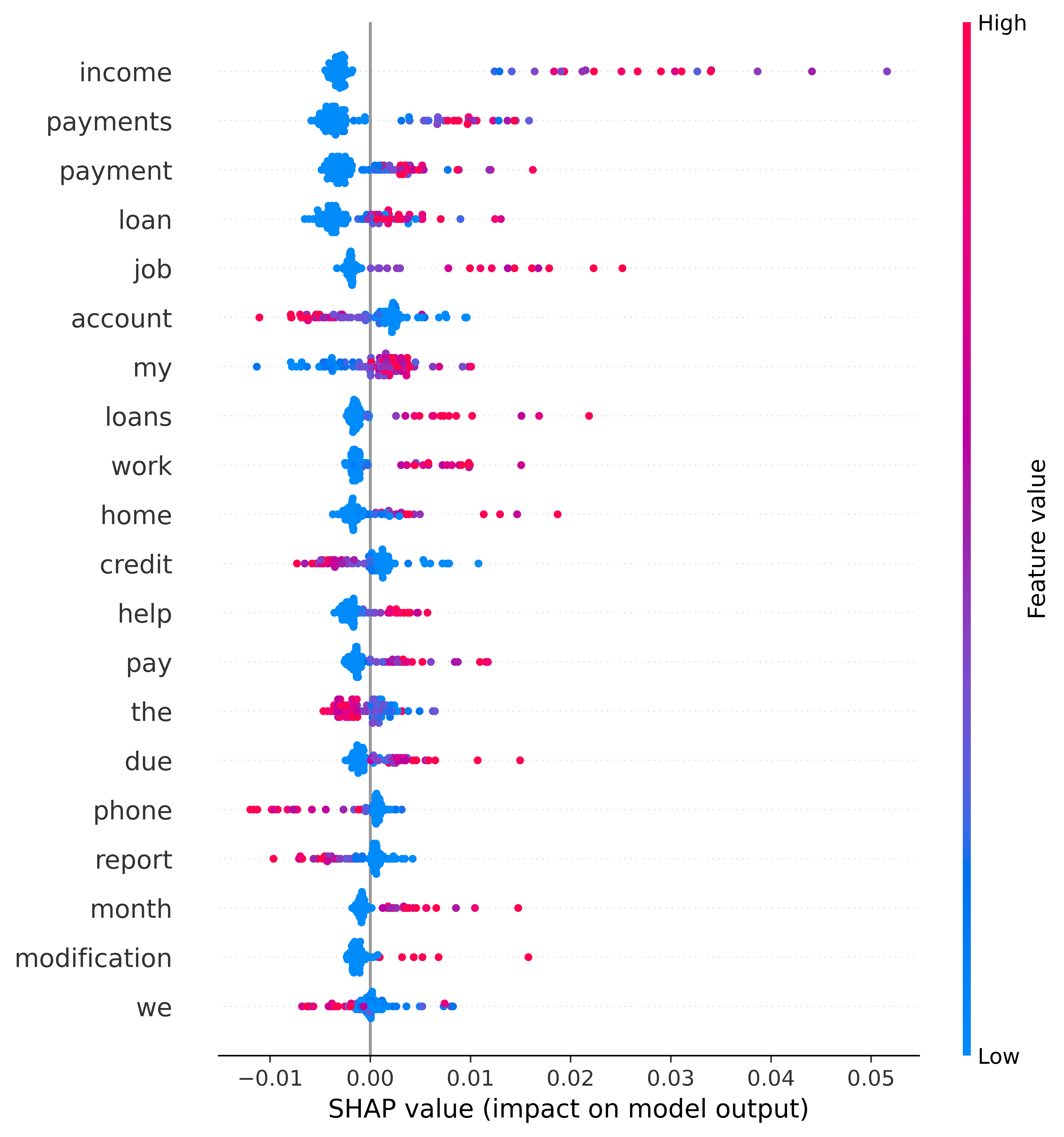}
    \end{subfigure}
     \hfill
\begin{subfigure}[b]{0.23\textwidth}
         \centering
    \includegraphics[width=\textwidth]{figures/train_explainer/1_train_explainer.png}
     \end{subfigure}
     \hfill
\begin{subfigure}[b]{0.223\textwidth}
         \centering
    \includegraphics[width=\textwidth]{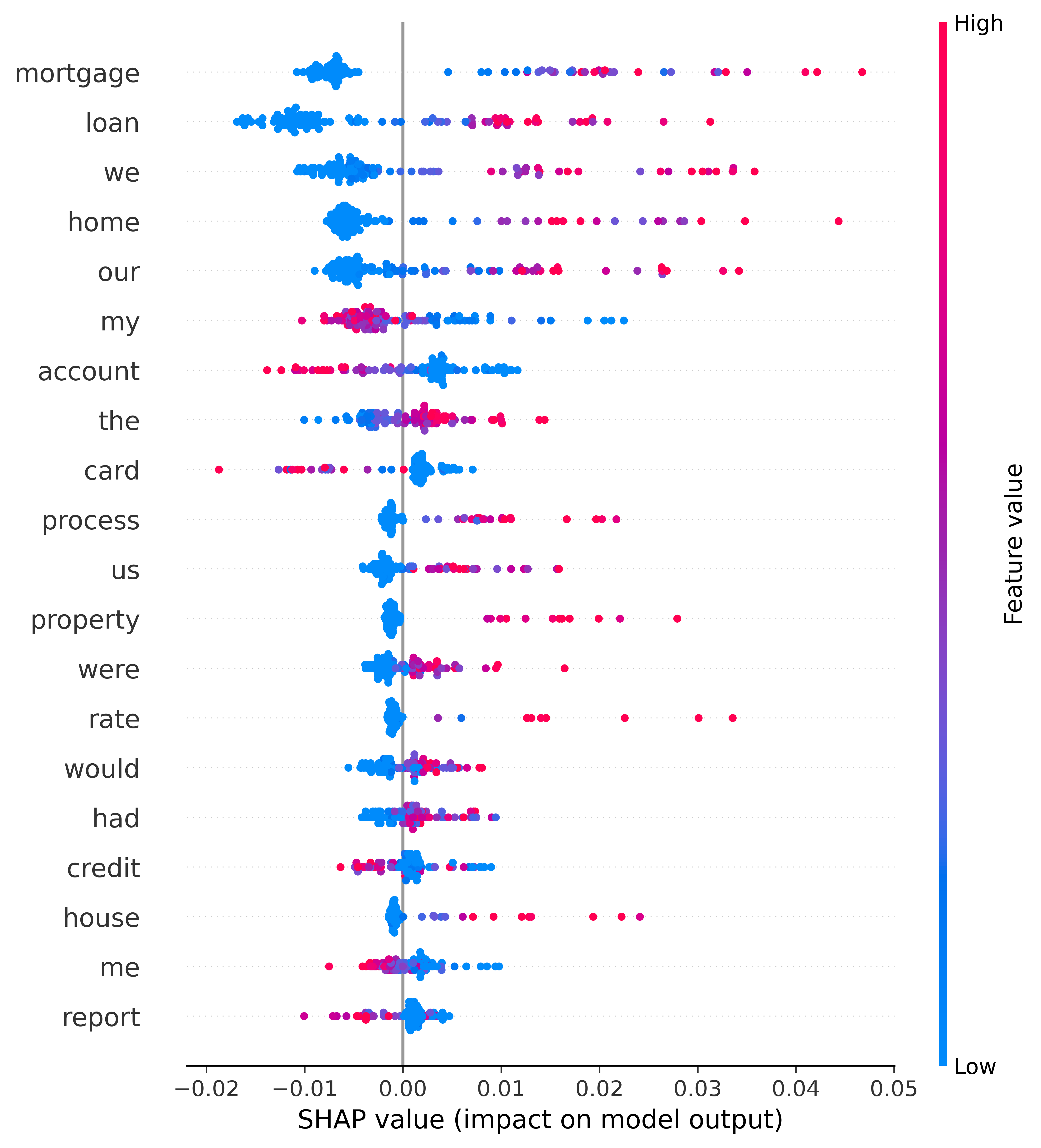}
     \end{subfigure}
     \hfill
\begin{subfigure}[b]{0.225\textwidth}
         \centering
    \includegraphics[width=\textwidth]{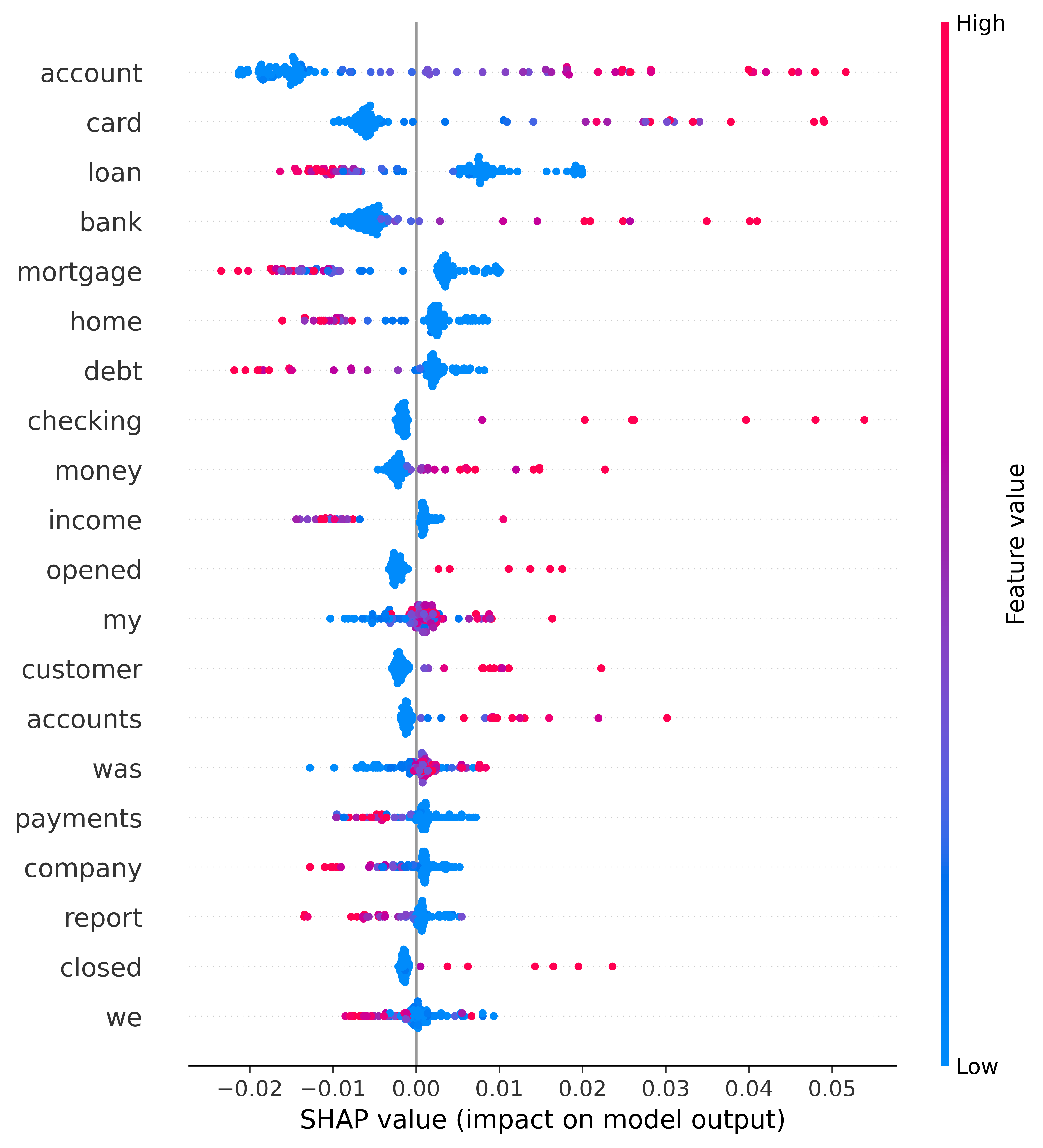}
     \end{subfigure}
     \hfill

\begin{subfigure}[b]{0.23\textwidth}
    \centering
    \includegraphics[width=\textwidth]{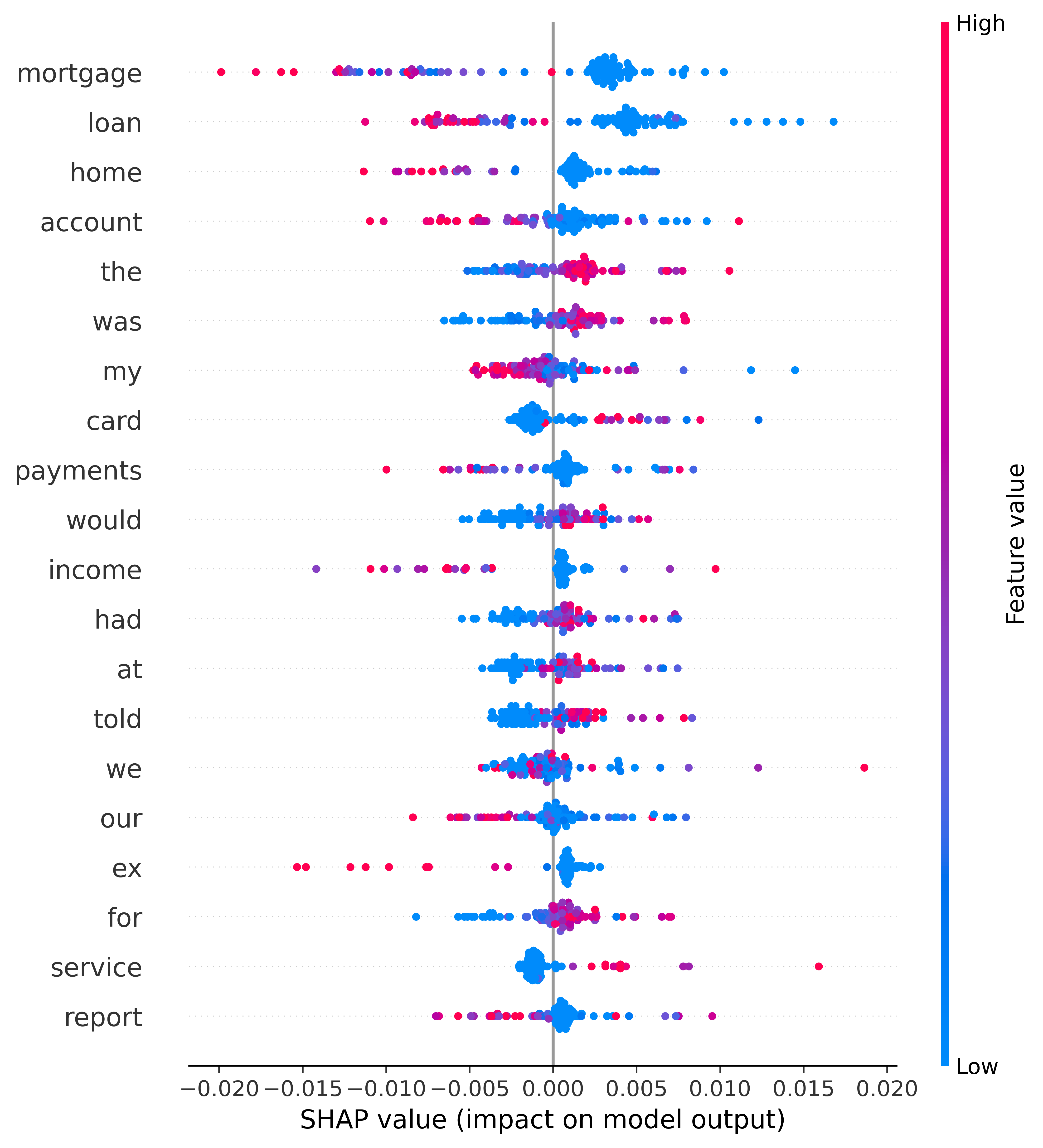}
    \end{subfigure}
     \hfill
\begin{subfigure}[b]{0.23\textwidth}
         \centering
    \includegraphics[width=\textwidth]{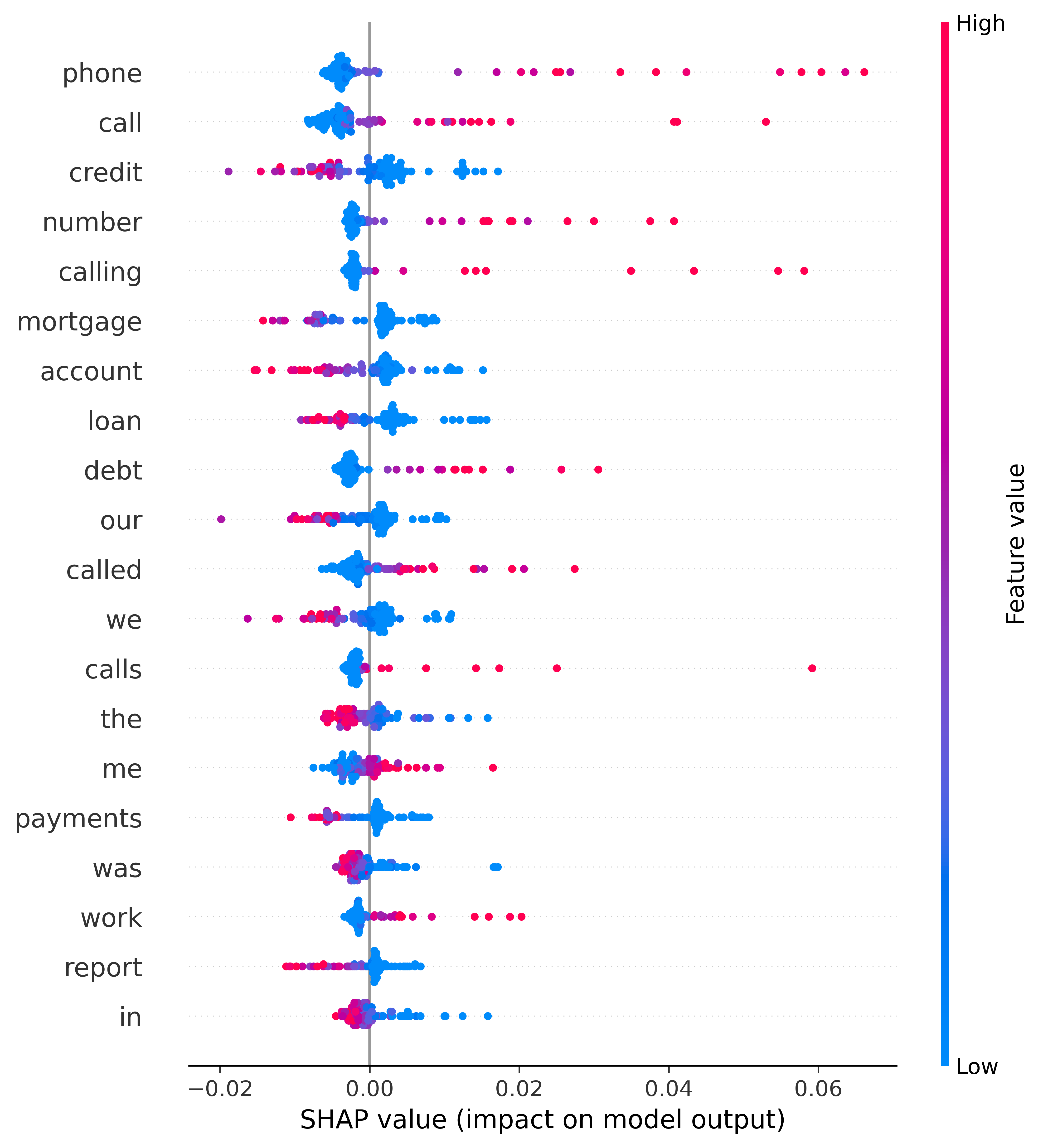}
     \end{subfigure}
     \hfill
\begin{subfigure}[b]{0.228\textwidth}
         \centering
    \includegraphics[width=\textwidth]{figures/train_explainer/6_train_explainer.png}
     \end{subfigure}
     \hfill
\begin{subfigure}[b]{0.235\textwidth}
         \centering
    \includegraphics[width=\textwidth]{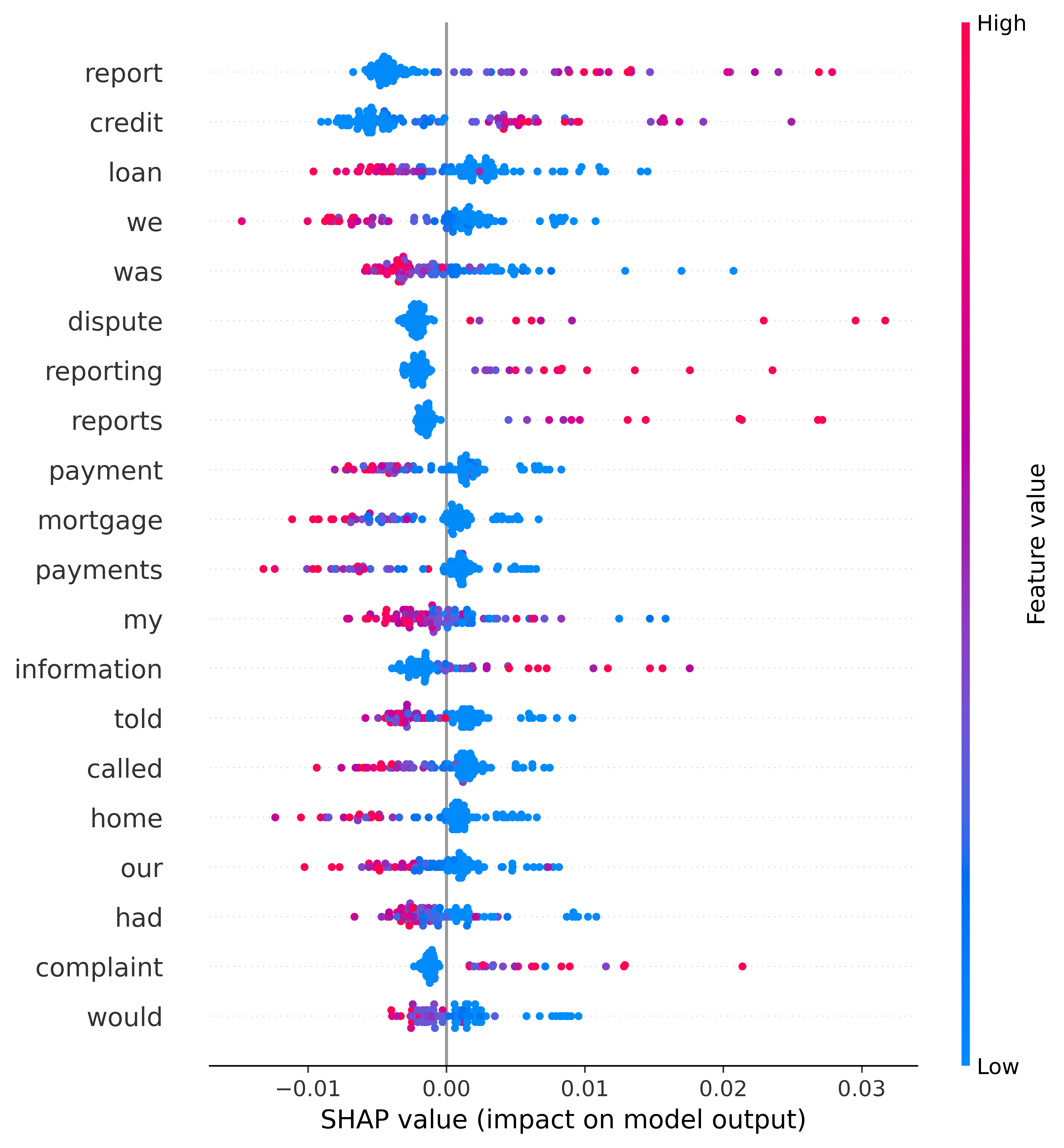}
     \end{subfigure}
     \hfill

\caption{SHAP scores for K-Clusters 0 to 7, numbered from top left to bottom right. The color bar corresponds to the raw values of the variables for each instance. If the variable for a particular word is high, it appears as a red dot, while low variable values appear as blue.}
\label{fig:SHAPs-scores}
\Description[SHAP score summaries for K-Clusters 0 to 7, numbered from top left to bottom right.]{SHAP scores for K-Clusters 0 to 7, numbered from top left to bottom right. The colour bar corresponds to the raw values of the variables for each instance. If the variable for a particular word is high, it appears as a red dot, while low variable values appear as blue.}
\end{figure}


\begin{table}[!htb]
\centering
\begin{tabular}{L{0.6\columnwidth}L{0.2\columnwidth}}
\toprule
\textbf{Data Field} & \textbf{Included} \\ \midrule
Date received &  \\
Product &  \\
Sub-product & \textit{Optional} \\
Issue &  \\
Sub-issue &  \\
Consumer complaint narrative & \textit{Optional} \\
Company public response & \textit{Optional} \\
Company &  \\
State &  \\
ZIP code &  \\
Tags & \textit{Optional} \\
Consumer consent provided? &  \\
Submitted via &  \\
Date sent to company &  \\
Company response to consumer &  \\
Timely response? &  \\
Customer disputed? & \textit{Optional} \\
Complaint ID &  \\ \bottomrule
\end{tabular}%
\caption{Database schema for CFPB data.}
\label{tab:cfpb-database-schema}
\Description[The data schema for the CFPB dataset. We use the "Consumer complaint narrative" in our study.]{The data schema for the CFPB dataset. We use the "Consumer complaint narrative" in our study.}
\end{table}


\begin{figure}[!htb]
    \vspace{-1.5em}
    \centering
    \includegraphics[width=0.45\textwidth, trim={0 0 1cm 1cm}, clip]{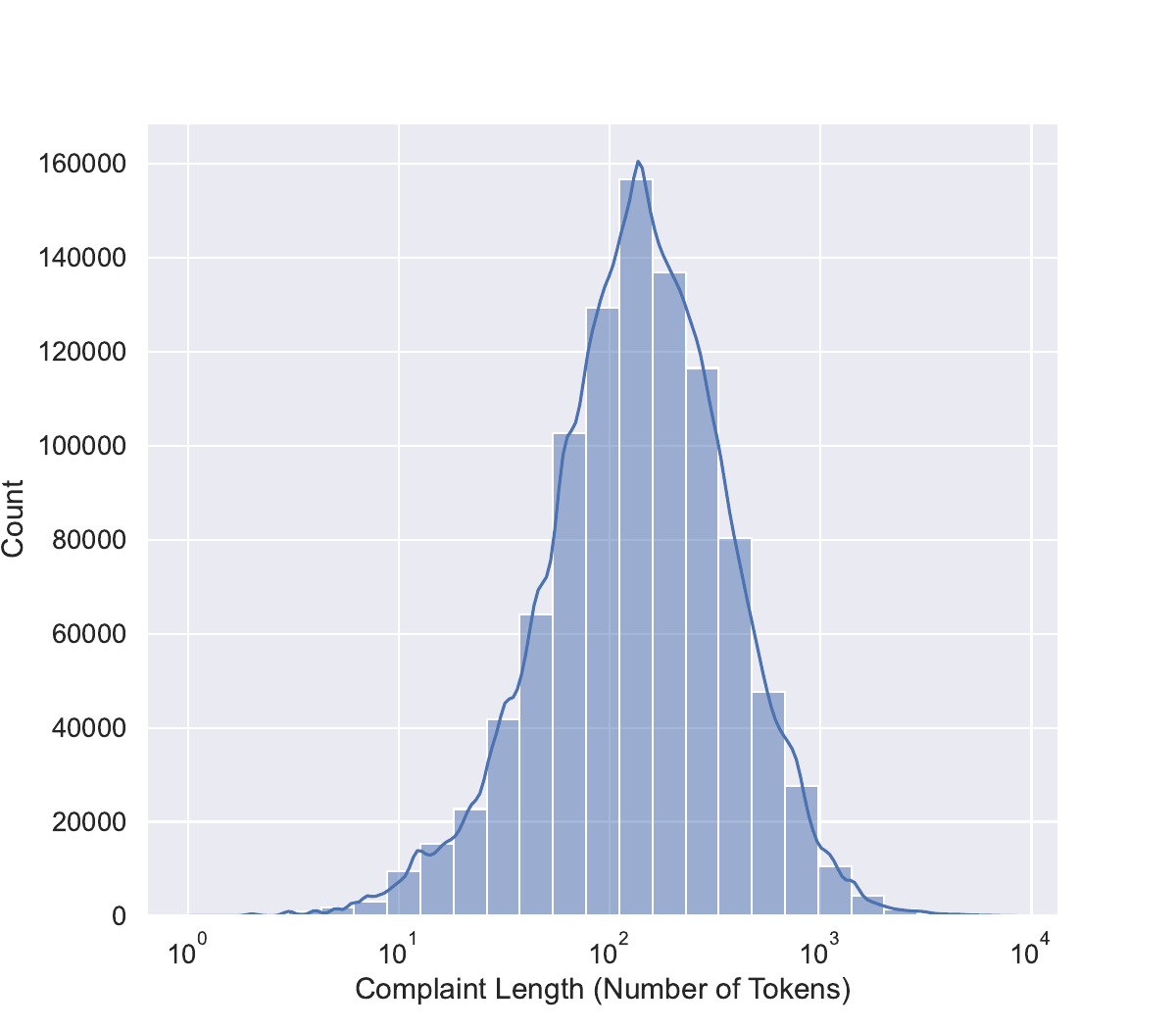}
    \caption{Distribution of complaint lengths within our CFPB Corpus}
    \label{fig:whole_corpus_lengths}
    \Description[A distribution of complaint lengths across the entire CFPB dataset we pulled. Complaint length is measured in tokens.]{A distribution of complaint lengths across the entire CFPB dataset we pulled. Complaint length is measured in tokens.}
\end{figure}

\section{Tools}
\label{sec:app-tools}
\begin{table}[!htb]
\centering
{ \begin{tabular}{L{0.5\columnwidth}L{0.2\columnwidth}}
\toprule
\textbf{Tool} & \textbf{Version} \\ \midrule
spacy & 3.1.3 \\
sentence-transformers & 2.2.2 \\
simcse & 0.4 \\
sklearn & 1.1.2 \\
shap & 0.39.0 \\ \bottomrule
\end{tabular}}
\caption{The tools and versions used in the instantiation and execution of our workflow.}
\label{tab:app-tools-versions}
\Description[Table displaying the relevant libraries and versions used in implementing our workflow for gathering complaints.]{Table displaying the relevant libraries and versions used in implementing our workflow for gathering complaints.}
\end{table}

\twocolumn[
    \section{Approach to qualitative coding}
    \label{sec:approach}
    \vspace{1.5em}
    \noindent\begin{minipage}{\textwidth}
        \centering
{\small \begin{tabular}{L{0.17\textwidth}L{0.43\textwidth}L{0.33\textwidth}}
\toprule
\textbf{Code category} & \textbf{Description of code category} & \textbf{Examples of low-level codes} \\ 
\midrule
\textit{Relationship Status} & Relationship status described in the complaint & Partner, Ex-Partner, Deceased Partner \\
\textit{Financial Product/Service} & Financial product(s) or service(s) contained in the complaint. & Loan: Car, Card: Credit, Account: {[}Unspecified{]} \\
\textit{Type of FA} & Type of financial abuse described here & Forging C's signature, Restrict access to C's account \\
\textit{Point of Discovery} & If otherwise unknown, the source of information on discovering FA & Online account: Bill due notice, Credit report \\
\textit{Method(s) of Resolution} & Steps to try and address, resolve or minimize the impact of FA & Submitted complaint to FSP, Closed account \\
\textit{Barriers to Help} & Barriers to deploying methods of resolution & Bank: ``Lost'' evidence, C: Lacks resources \\
\textit{Consequences of FA} & The negative consequences of FA to complainant and/or other & Substantial financial loss; FSP/CB not responding \\
\textit{Intimate Threat} & The presence of an intimate threat other than an intimate partner & Elder abuse, familial abuse, housemate abuse  \\
\bottomrule
\end{tabular}
        \Description[Table displaying the eight high-level coding categories used in a framework analysis, alongside a code description and examples.]{Table displaying the eight high-level coding categories used in a framework analysis, alongside a code description and examples. FA denotes financial abuse, C denotes complainant, FSP denotes financial service provider and CB denotes credit bureau. The coding categories are: Relationship Status, Financial Product/Service, Type of FA, Point of Discovery, Method(s) of Resolution, Barriers to Help, Consequences of FA, Intimate Threat.}}
        \captionof{table}{\edit{The eight high-level coding categories used in a framework analysis, alongside a code description and examples. FA denotes financial abuse, C denotes complainant, FSP denotes financial service provider and CB denotes credit bureau.}}
        \label{tab:code-cat}
    \end{minipage}
    \section{Corpus graphs}
    \vspace{1.5em}
    {\centering
        \includegraphics[width=0.8\textwidth]{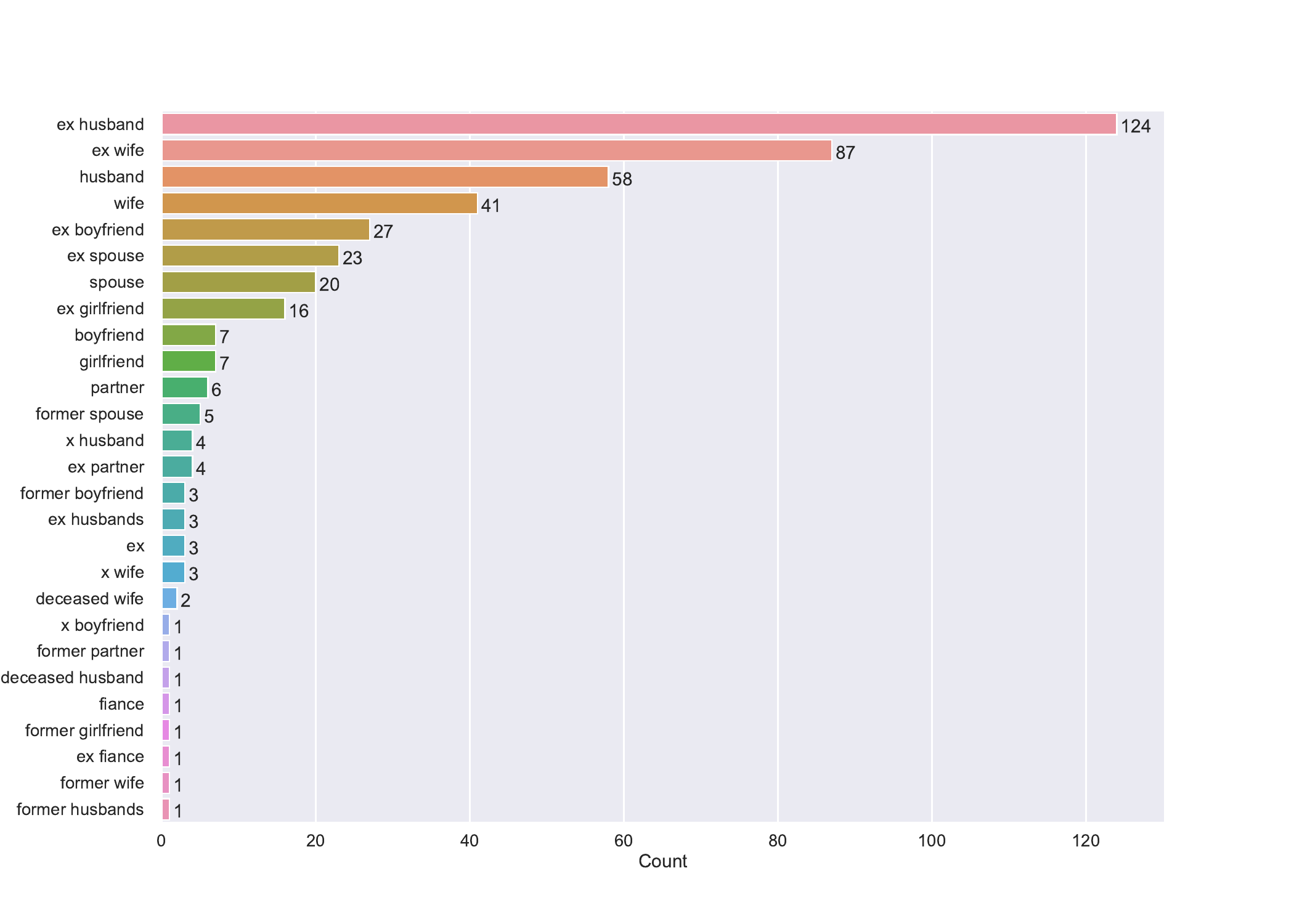}
        \captionof{figure}{Total count of intimate partner related keywords in our corpus}
        \label{fig:enter-label}
        \Description[Total count of different intimate partner related keywords in our corpus.]{Total count of intimate partner related keywords in our corpus. We note that "ex husband" is most frequently used with a count of 124, followed by "ex wife" with a frequency of 87. After that is "husband" with a frequency of 58. Then it is "wife" with a frequency of 41.}
    }
]



\end{document}